\documentclass[conference,compsoc]{IEEEtran}
\ifCLASSOPTIONcompsoc
  \usepackage[nocompress]{cite}
\else
  \usepackage{cite}
\fi
\ifCLASSINFOpdf
\else
\fi

\usepackage{cite}
\usepackage{algorithmic}
\usepackage{caption}
\usepackage{url}
\usepackage{xspace}
\usepackage{listings}
\usepackage{flushend}

\usepackage{amsmath}
\usepackage{amsfonts}
\usepackage{graphicx}
\usepackage{tikz}
\usepackage{algorithm}
\usepackage{pifont}
\usepackage{multirow}

\usepackage{tcolorbox}
\definecolor{mycolor}{rgb}{0.122, 0.435, 0.698}
\definecolor{codepurple}{rgb}{0.0,0,1.0} 
\lstdefinestyle{mystyle}{
   commentstyle=\color{codegreen},
   keywordstyle=\color{magenta},
   numberstyle=\footnotesize\color{codegray},
   stringstyle=\color{codepurple},
   basicstyle=\footnotesize\ttfamily,
   breakatwhitespace=false,
   breaklines=true,
   captionpos=b,
   keepspaces=true,
   numbers=left,
   numbersep=5pt,
   showspaces=false,
   showstringspaces=false,
   showtabs=false,
   tabsize=2,
   columns=fixed
}
\definecolor{darkgreen}{rgb}{0.0, 0.5, 0.0}
\definecolor{darkred}{rgb}{0.82, 0.1, 0.26}
\newcommand{\cmark}{\textcolor{darkgreen}{\ding{51}}}%
\newcommand{\xmark}{\textcolor{darkred}{\ding{55}}\ }%
\newcommand{\smart}{\textsc{AFL\-Smart}\xspace}
\newcommand{\fast}{\textsc{AFL\-fast}\xspace}
\newcommand{\vuzzer}{\textsc{VUzzer}\xspace}%
\begin{document}
 
\title{Smart Greybox Fuzzing\vspace{-0.5cm}}


\author{
\IEEEauthorblockN{Van-Thuan Pham\IEEEauthorrefmark{1}\quad
Marcel B\"{o}hme\IEEEauthorrefmark{2}\quad
Andrew E. Santosa\IEEEauthorrefmark{1}\\
Alexandru R\u{a}zvan C\u{a}ciulescu\IEEEauthorrefmark{3}\quad
Abhik Roychoudhury\IEEEauthorrefmark{1}}
\IEEEauthorblockA{\begin{tabular}{@{\quad\quad\ }rl@{}}\\[-0.2cm]
\IEEEauthorrefmark{1}National University of Singapore, Singapore &
thuanpv.nus@gmail.com, santosa\_1999@yahoo.com\\&abhik@comp.nus.edu.sg\\
\IEEEauthorrefmark{2}Monash University, Australia &
marcel.boehme@acm.org\\
\IEEEauthorrefmark{3}University Politehnica of Bucharest, Romania&
alexandru.razvan.c@gmail.com\vspace{-0.2cm}
\end{tabular}}
%
}



\maketitle

\begin{abstract}
Coverage-based greybox fuzzing (CGF) is one of the most successful methods for automated vulnerability detection.
Given a seed file (as a sequence of bits), CGF randomly flips, deletes or bits to generate new files. CGF iteratively constructs (and fuzzes) a seed corpus by retaining those generated files which enhance coverage. However, random bitflips are unlikely to produce valid files (or valid chunks in files), for applications processing complex file formats.

In this work, we introduce smart greybox fuzzing  (SGF) which leverages a high-level structural representation of the seed file to generate new files. We define innovative mutation operators that work on the virtual file structure rather than on the bit level which allows SGF to explore completely new input domains while maintaining file validity. We introduce a novel validity-based power schedule that enables SGF to spend more time generating files that are more likely to pass the parsing stage of the program, which can expose vulnerabilities much deeper in the processing logic.

Our evaluation demonstrates the effectiveness of SGF. On several libraries that parse structurally complex files, our tool \smart explores substantially more paths (up to 200\%) and exposes more vulnerabilities than baseline AFL. Our tool \smart has discovered 42 zero-day vulnerabilities in widely-used, well-tested tools and libraries; so far 17 CVEs were assigned.
\end{abstract}




\section{Introduction}
\label{sec:introduction}
\IEEEPARstart{C}{overage-based} greybox fuzzing (CGF) is a popular and effective approach
for software vulnerability detection. As opposed to blackbox approaches which
suffer from a lack of knowledge about the application, and whitebox approaches
which incur high overheads due to program analysis and constraint solving, greybox
approaches use lightweight code instrumentation. The American Fuzzy Lop (AFL) fuzzer \cite{afl18} and its
extensions \cite{boehme16coverage,boehme17aflgo,li17steelix,peng18fuzzing,stephens16driller,lemieux17fairfuzz,chen18angora} constitute the most widely-used embodiment of CGF.

CGF technology proceeds by input space exploration via mutation. Starting with
seed inputs, it mutates them using a pre-defined set of generic mutation
operators (such as bitflips). Control flows exercised by the mutated inputs
are then examined to determine whether they are sufficiently ``interesting''. The lightweight
program instrumentation helps the fuzzer make
this judgment on the novelty of the control flows. Subsequently, the mutated inputs
which are deemed sufficiently new are submitted for further investigation, at which
point they are mutated further to explore more inputs. The aim is to enhance
greater behavioral coverage, and to expose more vulnerabilities in a limited time budget.


One of the most significant and well-known limitations of CGF is its \emph{lack of input structure awareness}. The mutation operators of CGF work on the \emph{bit-level representation} of the seed file. Random bits are flipped, deleted, added, or copied from the same or from a different seed file. Yet, many security-critical applications and libraries will process highly structured inputs, such as image, audio, video, database, document, or spreadsheet files. Finding vulnerabilities effectively in applications processing such widely used formats is of imminent need. Mutations of the bit-level file representation are unlikely to effect any structural changes on the file that are necessary to effectively explore the vast yet sparse domain of valid program inputs. More likely than not arbitrary bit-level mutations of a valid file will result in an invalid file that is rejected by the program's parser before reaching the data processing portion of the program.

To tackle this problem, two main approaches have been proposed that are based on dictionaries \cite{aflGrammar} and dynamic taint analysis \cite{rawat17vuzzer}. Micha\l\ Zalewski, the creator of AFL, introduced the \emph{dictionary}, a lightweight technique to inject interesting byte sequences or tokens into the seed file during mutation at random locations. Zalewski's main concern \cite{dfchacker} was that a full support of input awareness might come at a cost of efficiency or usability, both of which are AFL's secret to success. 
AFL benefits tremendously from a dictionary when it needs to come up with magic numbers or chunk identifiers to explore new paths. Rawat et al.\cite{rawat17vuzzer} leverage dynamic taint analysis \cite{sp10taint} and control flow analysis to infer the locations and the types of the input data based on which their tool (\vuzzer) knows where and how to mutate the input effectively. However, both the dictionary and taint-based approaches do not solve our primary problem: to mutate the high-level structural representation of the file rather than its bit-level representation. For instance, neither a dictionary nor an inferred program feature help in adding or deleting complete chunks from a file.

In contrast to CGF, smart blackbox fuzzers \cite{peach18,holler12fuzzing} are already input-structure aware and leverage a model of the file format to construct new valid files from existing valid files. For instance, Peach \cite{peach18} uses an input model to disassemble valid files and to reassemble them to new valid files, to delete chunks, and to modify important data values. LangFuzz \cite{holler12fuzzing} leverages a context-free grammar for JavaScript (JS) to extract code fragments from JS files and to reassemble them to new JS files. However, awareness of input structure alone is insufficient and the coverage-feedback of a greybox fuzzer is urgently needed -- as shown by our experiments with Peach. In our experiments Peach performs much worse even than AFL, our baseline greybox fuzzer. Our detailed investigation revealed that Peach does not reuse the generated inputs that improve coverage for further test input generation. For instance, if Peach generated a WAV-file with a different (interesting) number of channels, that file could not be used to generate further WAV-files with the newly discovered program behaviour. Without coverage-feedback interesting files will not be retained for further fuzzing. On the other hand, retaining all generated files would hardly be economical.

In this paper, we introduce \emph{smart greybox fuzzing} (SGF)---which leverages a high-level structural representation of the seed file to generate new files---and investigate the impact on fuzzer efficiency and usability. We define innovative mutation operators that work on the virtual structure of the file rather than on the bit level. These \emph{structural mutation operators} allow SGF to explore completely new input domains while maintaining the validity of the generated files. We introduce a novel \emph{validity-based power schedule} that assigns more energy to seeds with a higher degree of validity and enables SGF to spend more time generating files that are more likely to pass the parsing stage of the program to discover vulnerabilities deep in the processing logic of the program.

We implement \smart, a robust yet efficient and easy-to-use \emph{smart} greybox fuzzer based on AFL, a popular and very successful CGF. \smart integrates the input-structure component of Peach with the coverage-feedback component of AFL. Hence, in our evaluation we compare against both as baseline techniques. Our evaluation demonstrates that \smart, within a given time limit of 24 hours, can double the zero-day bugs found. \smart discovers 33 bugs (8 CVEs assigned) while the baseline (AFL and its extension \fast \cite{boehme16coverage}) can detect only 16 bugs, in large, widely-used, and well-fuzzed open-source software projects, such as FFmpeg, LibAV, LibPNG, Wavpack, OpenJPEG and Binutils.
\smart also significantly improves the path coverage up to 200\% compared to the baseline. \smart also outperforms \vuzzer~\cite{rawat17vuzzer} on its benchmarks; \smart discovers seven (7) bugs which \vuzzer could not find in another set of popular open-source programs, such as \emph{tcpdump}, \emph{tcptrace} and \emph{gif2png}. Moreover, in a 1-week bug hunting campaign for FFmpeg, \smart discovers nine (9) more zero-day bugs (9 CVEs assigned). 
Its effectiveness comes with negligible overhead -- with our optimization of \emph{deferred cracking} \smart achieves execution speeds which are similar to AFL.

In our experience with \smart, the time spent writing a file format specification is outweighed by the tremendous improvement in behavioral coverage and the number of bugs exposed. One of us spent five working days to develop 10 file format specifications (as Peach Pits \cite{peach18}) which were used to fuzz all 16 subject programs. Hence, once developed, file format specifications can be reused across programs as well as for different versions of the same program. 

In summary, the main contribution of our work is to make greybox fuzzing input format-aware.
Given an input format specification (e.g., a Peach Pit \cite{peach18}), our \emph{smart greybox fuzzer} derives a structural representation of the seed file, called virtual structure, and leverages our novel smart mutation operators to modify the virtual file structure in addition to the file's bit sequence during the generation of new input files.
We propose smart mutation operators, which are likely to preserve the satisfaction w.r.t. a file format specification. 
During the greybox fuzzing search, our tool \smart measures the degree of validity of the inputs produced with respect to the file format specification. It prioritizes valid inputs over invalid ones, by
enabling the fuzzer to explore more mutations of a valid file as opposed to an invalid one. 
As a result, our smart fuzzer largely explores the restricted space of inputs which are valid as per the file format specification, and attempts to locate vulnerabilities in the file processing logic
by running inputs in this restricted space.
We conduct extensive evaluation on well-tested subjects processing complex file formats such as PNG and WAV. Our experiments demonstrate that the smart mutation
operators and the validity-based power schedule introduced by us, increases the effectiveness of fuzzing both in terms of path coverage and vulnerabilities found within a time limit of 24 hours.
These results also demonstrate that the additional effectiveness in our smart fuzzer \smart is {\em not} achieved by sacrificing the efficiency of greybox fuzzing and AFL.

\section{Motivating Example}
\label{sec:motivating}

\subsection{The WAVE File Format}
Most file systems store information as a long string of zeros and ones---a file. It is the task of the program to make sense of this sequence of bits, i.e., to parse the file, and to extract the relevant information. This information is often structured in a hierarchical manner which requires the file to contain additional structural information. The structure of files of the same type is defined in a file format. Adherence to the file format allows the same file to be processed by different programs.

\begin{figure}[h]\footnotesize
\vspace{-0.1cm}
\begin{tabular}{@{}cllp{2.9cm}@{}}
\textbf{Chunk Type} & \textbf{Field} & \textbf{Length} & \textbf{Contents}\\\hline
\multirow{4}{*}{RIFF} & \texttt{ckID} & 4 & Chunk ID: \texttt{RIFF}\\
& \texttt{cksize} & 4 & Chunk size: 4+$n$\\
& \texttt{WAVEID} & 4 & WAVE id: \texttt{WAVE}\\
& chunks &$n$& Chunks containing format information and sampled data\\\hline
\multirow{6}{*}{fmt} & \texttt{ckID} & 4 & Chunk ID: \texttt{fmt}\\
& \texttt{cksize} & 4 & Chunk size: 16, 18 or 40\\
&\texttt{wFormatTag}&	2&	Format code\\
&\texttt{nChannels}&	2&	Number of interleaved channels\\
&\texttt{nSamplesPerSec}&	4&	Sampling rate (blocks per second)\\
&	\ldots \\\hline
\multicolumn{4}{@{}l@{}}{\emph{Optional chunks (fact chunk, cue chunk, playlist chunk, \ldots)}\rule{0pt}{2.6ex}}\\[2pt] \hline
\multirow{4}{*}{data}&\texttt{ckID}&	4&	Chunk ID: \texttt{data}\\
&\texttt{cksize}&	4&	Chunk size: $n$\\
&\texttt{sampled data}&$n$&	Samples\\
&\texttt{pad byte}&	0 or 1&	Padding byte if $n$ is odd\\\hline
\end{tabular}
\vspace{-0.1cm}
\caption{An excerpt of the WAVE file format (from Ref.~\cite{WAVE})}
\label{tab:wave}
\end{figure}

WAVE files (\texttt{*.wav}) contain audio information and can be processed by various media players and editors. A WAVE file consists of \emph{chunks} (see Figure~\ref{tab:wave}). Each chunk consists of chunk identifier, chunk length and chunk data. Chunks are structured in a hierarchical manner. The root chunk requires the first four bytes of the file to spell (in unicode) \texttt{RIFF} followed by four bytes specifying the total size $n$ of the children chunks plus four. The next four bytes must spell (in unicode) \texttt{WAVE}. The remainder of a WAVE file contains the children chunks, the mandatory \texttt{fmt} chunk, several optional chunks, and the \texttt{data} chunk. The \texttt{data} chunk itself is subject to further structural constraints.

We can clearly see that a WAVE file embeds audio information and meta-data in a hierarchical chunk structure. The WAVE file format governs all WAVE files and allows for efficient and systematic parsing of the audio information.

\subsection{The Anatomy of a Vulnerability in a Popular Audio Compression Library}
In the following, we discuss a vulnerability that our smart greybox fuzzer \smart found in WavPack \cite{wavPack}, a popular audio compression library that is used by many well-known media players and editors such as Winamp, VLC Media Player, and Adobe Audition. In our experiments, the same vulnerability could not be found by traditional greybox fuzzers such as AFL \cite{afl18} or \fast \cite{boehme16coverage}.

The discovered vulnerability (CVE-2018-10536) is a \emph{buffer overwrite} in the WAVE-parser component of WavPack.
To construct an exploit, a \emph{WAVE file} with more than one format chunks needs to be crafted that satisfies several complex structural conditions. The WAVE file contains the mandatory \texttt{riff}, \texttt{fmt}, and \texttt{data} chunks, plus an additional \texttt{fmt} chunk placed right after the first \emph{fmt} chunk.
The first \texttt{fmt} chunk specifies IEEE 754 32-bits (single-precision) floating point (\emph{IEEE float\/}) as the waveform data format (i.e., fmt.\texttt{wFormatTag}$=3$) and passes all sanity checks. The second \texttt{fmt} chunk specifies PCM as the waveform data format, one channel, one bit per sample, and one block align (i.e., fmt.\texttt{wFormatTag}$=1$, fmt.\texttt{nChannels}$=1$, fmt.\texttt{nBlockAlign}=1, and fmt.\texttt{wBitsPerSample}$=1$).

\begin{figure}[h]
\begin{tabular}{@{}r@{ \ }l}
\tiny{1}&\footnotesize \texttt{\textbf{else if} (!strncmp (chunk\_header.ckID, "fmt ", 4))\{}\\
\tiny{2}&\footnotesize \texttt{DoReadFile (infile, \&WaveHeader, \ldots)}\\
\tiny{3}&\footnotesize \texttt{format = WaveHeader.FormatTag;}\\
\tiny{4}&\footnotesize \texttt{config->bits\_per\_sample = WaveHeader.BitsPerSample;}\\[0.05cm]
\tiny{5}&\footnotesize \emph{// Sanity checks}\\
\tiny{6}&\footnotesize \texttt{if (format == 3 \&\& config->bits\_per\_sample != 32)}\\
\tiny{7}&\footnotesize \texttt{\quad supported = FALSE;}\\
\tiny{8}&\footnotesize \texttt{if (WaveHeader.BlockAlign / WaveHeader.NumChannels}\\
        &\footnotesize \texttt{\quad\quad\quad < (config->bits\_per\_sample + 7) / 8})\\
\tiny{9}&\footnotesize \texttt{\quad supported = FALSE;}\\
\tiny{10}&\footnotesize \texttt{if (!supported) exit();}\\[0.05cm]
\tiny{11}&\footnotesize \texttt{if (format==3) config->float\_norm\_exp=CONFIG\_FLOAT;}\\
\tiny{12}&\footnotesize \texttt{\ldots }

\end{tabular}
\caption{Sketching \texttt{cli/riff.c} @ revision \texttt{0a72951}}
\label{fig:vuln2}
\end{figure}

The first \texttt{fmt} chunk configures WavPack to read the data in \emph{IEEE float} format, which requires certain constraints to be satisfied, e.g., on the number of bits per sample (Lines 6--10). The second \texttt{fmt} chunk allows to override certain values, e.g., the number of bits per sample, while maintaining the \emph{IEEE float} format configuration. More specifically, the \texttt{fmt}-handling code is shown in Figure~\ref{fig:vuln2}. The first \texttt{fmt} chunk is parsed as format 3 (IEEE float), 32 bits per sample, 1 channel, and 4 block align (Lines 2--4). The configuration passes all sanity checks for an \emph{IEEE float} format (Lines~6--10), and sets the global configuration accordingly (Line~11).  The second \texttt{fmt} chunk is parsed as format 1 (PCM), 1 bits per sample, 1 channel, and 1 block align (Lines 2--4). The new configuration would be valid if WavPack had not maintained \emph{IEEE float} as the waveform data and had reset \texttt{float\_norm\_exp}. However, it does maintain \emph{IEEE float} and thus allows an invalid configuration that would otherwise not pass the sanity checks
which finally leads to a buffer overwrite that can be controlled by the attacker.

The vulnerability was patched by aborting when the \texttt{*.wav} file contains more than one \texttt{fmt} chunk. A similar vulnerability (CVE-2018-10537) was discovered and patched for \texttt{*.w64} (WAVE64) files. 

\subsection{Difficulties of Traditional Greybox Fuzzing}

\definecolor{shadecolor}{RGB}{200,200,200}
 \renewcommand{\algorithmicrequire}{\textbf{Input:}}
\renewcommand{\algorithmicensure}{\textbf{Output:}}
\begin{algorithm}[h]
\caption{Coverage-based Greybox Fuzzing}
\label{alg:greybox}
\begin{algorithmic}[1]
    \REQUIRE Seed Corpus $S$
    \REPEAT \label{line:mainloop}
      \STATE \hspace{-0.14cm}\colorbox{white}{\parbox{\dimexpr\columnwidth-10.5\fboxsep\relax}{$s = $ \textsc{chooseNext}$(S)$ \hfill \emph{// Search Strategy}}} \label{line:chooseNext}
      \STATE \vspace{-0.1cm}\hspace{-0.14cm}\colorbox{white}{\parbox{\dimexpr\columnwidth-10.5\fboxsep\relax}{$p =$ \textsc{assignEnergy}$(s)$ \hfill \emph{// Power Schedule}}} \label{line:assignEnergy}
      \FOR{$i$ from 1 to $p$}
        \STATE \hspace{-0.14cm}\colorbox{shadecolor}{\parbox{\dimexpr\columnwidth-15.5\fboxsep\relax}{$s' =$ \textsc{mutate\_input}$(s)$}} \label{line:mutate}
        \IF{$s'$ crashes}
          \STATE add $s'$ to $S_\text{\xmark}$
        \ELSIF{\textsc{isInteresting}$(s')$}
          \STATE add $s'$ to $S$
        \ENDIF
      \ENDFOR
    \UNTIL{\emph{timeout} reached or \emph{abort}-signal} \label{line:endmainloop}
    \ENSURE Crashing Inputs $S_\text{\xmark}$
\end{algorithmic}
\end{algorithm}

We use these vulnerabilities to illustrate the shortcomings of traditional greybox fuzzing.
Algorithm~\ref{alg:greybox}, which is extracted from \cite{boehme16coverage}, shows the general greybox fuzzing loop. The fuzzer is provided with a initial set of program inputs, called \emph{seed corpus}. In our example, this could be a set of WAVE files that we know to be valid. The greybox fuzzer mutates these seed inputs  in a continuous loop to generate new inputs. Any new input that increases the coverage is added to the seed corpus. A well-known and very successful coverage-based greybox fuzzer is American Fuzzy Lop (AFL) \cite{afl18}.

\emph{Guidance}. A coverage-based greybox fuzzer is guided by a search strategy and a power schedule. The \emph{search strategy} decides the order in which seeds are chosen from the seed corpus, and is implemented in \textsc{chooseNext} (Line~\ref{line:chooseNext}). The \emph{power schedule} decides a seed's energy, i.e., how many inputs are generated by fuzzing the seed, and is implemented in \textsc{assignEnergy} (Line~\ref{line:assignEnergy}). For instance, AFL spends more energy fuzzing seeds that are small and execute quickly.

\emph{Bit-level mutation}. Traditional greybox fuzzers are unaware of the input structure. In order to generate new inputs, a seed is modified according to pre-defined mutation operators.
 A \emph{mutation operator} is a transformation rule. For instance, a bit-flip operator turns a zero into a one, and vice versa. Given a seed input, a mutation site is randomly chosen in the seed input and a mutation operator applied to generate a new test input. In Algorithm~\ref{alg:greybox}, the method \textsc{mutate\_input} implements the input generation by seed mutation. These mutation operators are specified on the \emph{bit-level}. For instance, AFL has several deletion operators, all of which delete a contiguous, fixed-length sequence of bits in the seed file. AFL also has several addition operators, for instance to add a sequence of only zero's or one's, a random sequence of bits, or to copy a sequence of bits within the file.
For our motivating example, Figure~\ref{tab:wavefile} shows the first 72 bytes of a canonical WAVE file. 
To expose CVE-2018-10536, a second \emph{valid} \texttt{fmt} chunk must be added in-between the existing \texttt{fmt} and \texttt{data} chunks. Clearly, it is extremely unlikely for AFL to apply a sequence of bit-level mutation operators to the file that result in the insertion of such additional, valid chunks.

\begin{figure}\footnotesize\centering
\begin{tabular}{@{}l||rl@{}}
\textbf{Stored Bits} & \textbf{Information} & \textbf{Description}\\
\texttt{52 49 46 46} & \texttt{ R \ I \ F \ F}&RIFF.\texttt{ckID}\\
\texttt{24 08 00 00} & 2084 & RIFF.\texttt{cksize}\\
\texttt{57 41 56 45} & \texttt{ W \ A \  V \ E} & RIFF.\texttt{WAVEID}\\\cline{2-3}
\texttt{66 6d 74 20} & \texttt{f \ m \ t \ \textvisiblespace} & fmt.\texttt{ckID} \\
\texttt{10 00 00 00} & 16& fmt.\texttt{cksize}\\
\texttt{01 00 02 00} & 1 \quad\quad 2 & fmt.\texttt{wFormatTag} (1=PCM) \&\\
\texttt{                 } &    \quad\quad    & fmt.\texttt{nChannels}\\
\texttt{22 56 00 00} & 22050 & fmt.\texttt{nSamplesPerSec}\\
\texttt{88 58 01 00} & 88200 & fmt.\texttt{nAvgBytesPerSec}\\
\texttt{04 00 10 00} & 4 \ \ \ \quad 16& fmt.\texttt{nBlockAlign} \&\\
\texttt{                 } &    \ \ \ \quad     & fmt.\texttt{wBitsPerSample}\\\cline{2-3}
\texttt{64 61 74 61} & \texttt{d \ a \ t \ a} & data.\texttt{ckID}\\
\texttt{00 08 00 00} & 2048 & data.\texttt{cksize}\\
\texttt{00 00 00 00} & sound data 1 &left and right channel\\
\texttt{24 17 1e f3} & sound data 2 &left and right channel\\
\texttt{3c 13 3c 14} & sound data 3 &left and right channel\\
\texttt{16 f9 18 f9} & sound data 4 &left and right channel\\
\texttt{34 e7 23 a6} & sound data 5 &left and right channel\\
\texttt{3c f2 24 f2} & sound data 6 &left and right channel\\
\texttt{11 ce 1a 0d} & sound data 7 &left and right channel\\
\multicolumn{2}{@{}l}{\ldots}
\end{tabular}
\caption{Canonical WAVE file  (from Ref.~\cite{WAVE})}
\label{tab:wavefile}
\end{figure}

\emph{Dictionary}. To better facilitate the fuzzing of structured files, many greybox fuzzers, including AFL, allow to specify a list of interesting byte sequences, called dictionary. In our motivating example, such byte sequences could be words, such as \texttt{RIFF}, \texttt{fmt}, and \texttt{data} in unicode, or common values, such as 22050 and 88200 in hexadecimal. However, a dictionary will not contribute much to the complex task of constructing a valid chunk that is inserted right at the joint boundary of two other chunks. 

\section{Smart Greybox Fuzzing}
\label{sec:grammar}

Smart greybox fuzzing (SGF) is more effective than both, smart blackbox fuzzing and traditional greybox fuzzing. Unlike traditional greybox fuzzing, SGF allows to penetrate deeply into a program that takes highly-structured inputs without getting stuck in the program's parser code. Unlike smart blackbox fuzzing, SGF leverages coverage-information to explore the program's behavior more efficiently.

\subsection{Virtual Structure}
The effectiveness of SGF comes from the careful design of its smart mutation operators. First, these operators should fully leverage the structural information extracted from the seed inputs to apply higher-order manipulations at both the chunk level and the bit level.
Second, they should be unified operators to support all chunk-based file formats (e.g., MP3, ELF, PNG, JPEG, WAV, AVI, PCAP).
Last but not the least, all these operators must be lightweight so that we can retain the efficiency of greybox fuzzing.

\begin{figure}[h]\centering
\includegraphics[width=0.8\columnwidth]{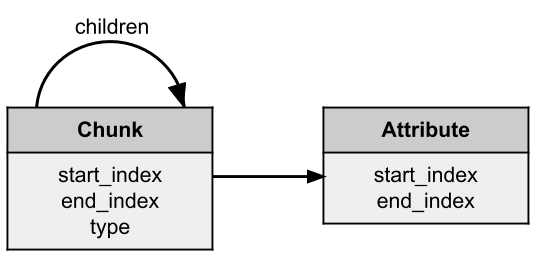}
\caption{Virtual structure used by \smart}
\label{fig:structure}
\end{figure}

To implement these three design principles, we introduce a new lightweight yet generic data structure namely \emph{virtual structure} which can facilitate the structural mutation operators.
Each input file can be represented as a (parse) tree. The nodes of this tree are called chunks or attributes, with the {\em chunks} being the internal nodes of the tree and the {\em attributes} being the leaf nodes of the tree.

A \emph{chunk} is a contiguous sequence of bytes in the file. There is a \emph{root chunk} spanning the entire file. As visualized in Fig.~\ref{fig:structure}, each chunk has a \emph{start- and an end-index} representing the start and end of the byte sequence in the file, and a \emph{type} representing the distinction to other chunks (e.g., an \texttt{fmt} chunk is different from a \texttt{data} chunk in the WAVE file format). Each chunk can have zero or more chunks as \emph{children} and zero or more attributes. An \emph{attribute} represents important data in the file that is not structurally relevant, for instance \texttt{wFormatTag} in the \texttt{fmt} chunk of a WAVE file. 

\lstset{%
   		framexleftmargin=0mm,
   		frame=single,
   		numbers=none,
		resetmargins=true,
   		basicstyle=\ttfamily\selectfont\footnotesize,
   	  frameround= tttt,
   	  breaklines = true,
   	  tabsize=1,
   	  captionpos=b,
   	  language=C,
   	  morekeywords={Bug}}
   {\footnotesize
\begin{lstlisting}[frame=single, caption=WAVE Peach Pit File Format Specification, label=lst:wav]
<DataModel name="Chunk">
  <String name="ckID" length="4"/>
  <Number name="cksize" size="32" >
    <Relation type="size" of="Data"/>
  </Number>
  <Blob name="Data"/>
  <Padding alignment="16"/>
</DataModel>
<DataModel name="ChunkFmt" ref="Chunk">
   <String name="ckID" value="fmt "/>
   <Block name="Data">
      <Number name="wFormatTag" size="16"/>
      <Number name="nChannels" size="16"/>
      <Number name="nSampleRate" size="32"/>
      <Number name="nAvgBytesPerSec" size="32"/>
      <Number name="nBlockAlign" size="16" />
      <Number name="nBitsPerSample" size="16"/>
   </Block>
</DataModel>
...
<DataModel name="Wav" ref="Chunk">
  <String name="ckID" value="RIFF"/>
  <String name="WAVE" value="WAVE"/>
  <Choice name="Chunks" maxOccurs="30000">
    <Block name="FmtChunk" ref="ChunkFmt"/>
    ...
    <Block name="DataChunk" ref="ChunkData"/>
  </Choice>
</DataModel>
\end{lstlisting}
}

As an example, the canonical WAVE file in Figure~\ref{tab:wavefile} has the following virtual structure. The root chunk has start and end index $\{0,2083\}$. The root chunk (\texttt{riff}) has three attributes, namely \texttt{ckID}, \texttt{cksize}, and \texttt{WAVEID}, and two children with indices $\{12,35\}$ and $\{36,2083\}$, respectively. The first child \texttt{fmt} has eight attributes namely \texttt{ckID}, \texttt{cksize}, \texttt{wFormatTag}, \texttt{nChannels}, \texttt{nSamplesPerSec}, \texttt{nAvgBytesPerSec}, \texttt{nBlockAlign}, and \texttt{wBitsPerSample}. 

To construct the virtual structure, a file format specification and a parser is required. Given the specification and the file, the parser constructs the virtual structure.
For example, Peach \cite{peach18} has a robust parser component called \emph{File Cracker}. Given an input file and the file format specification, called Peach Pit, our extension of the File Cracker precisely parses and decomposes the file into chunks and attributes and provides the boundary indices and type information. Listing~\ref{lst:wav} shows a snippet of the Peach Pit for the WAV file format. In this specification, we can specify the order, type, and structure of chunks and attributes in a valid WAV file. In Section~\ref{sec:specs} we explain how this specification can be constructed.

\subsection{Smart Mutation Operators}

Based on this virtual input structure, we define three generic structural mutation operators -- \emph{smart deletion}, \emph{smart addition} and \emph{smart splicing}.

\begin{figure}[h]\centering
\vspace{-0.2cm}
\includegraphics[width=0.9\columnwidth]{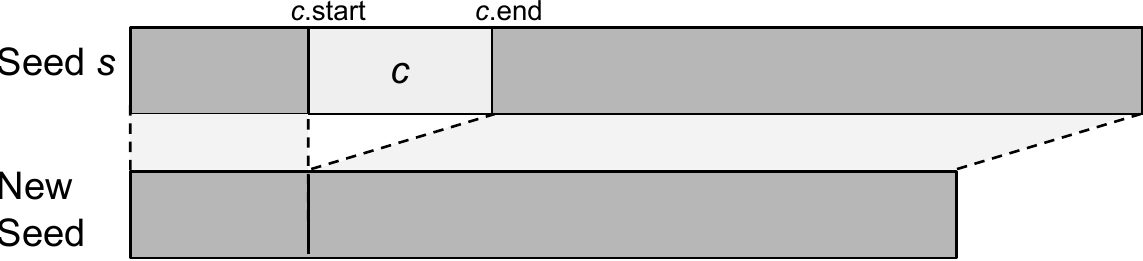}
\vspace{-0.2cm}
\end{figure}
\textbf{Smart Deletion}.
Given a seed file $s$, choose an arbitrary chunk $c$ and delete it.
The SGF copies the bytes following the end-index of the chosen chunk $c$ to the start-index of $c$, revises the indices of all affected chunks accordingly. 
For instance, to delete the \texttt{fmt}-chunk in our canonical WAVE file, the stored bits in the index range $[36,2083]$ are memcpy'd to index $12$. The indices in the virtual structure of the new WAVE file are revised. For instance, the \texttt{riff}-chunk's end index is revised to $2048$.

\begin{figure}[h]\centering
\vspace{-0.2cm}
\includegraphics[width=0.9\columnwidth]{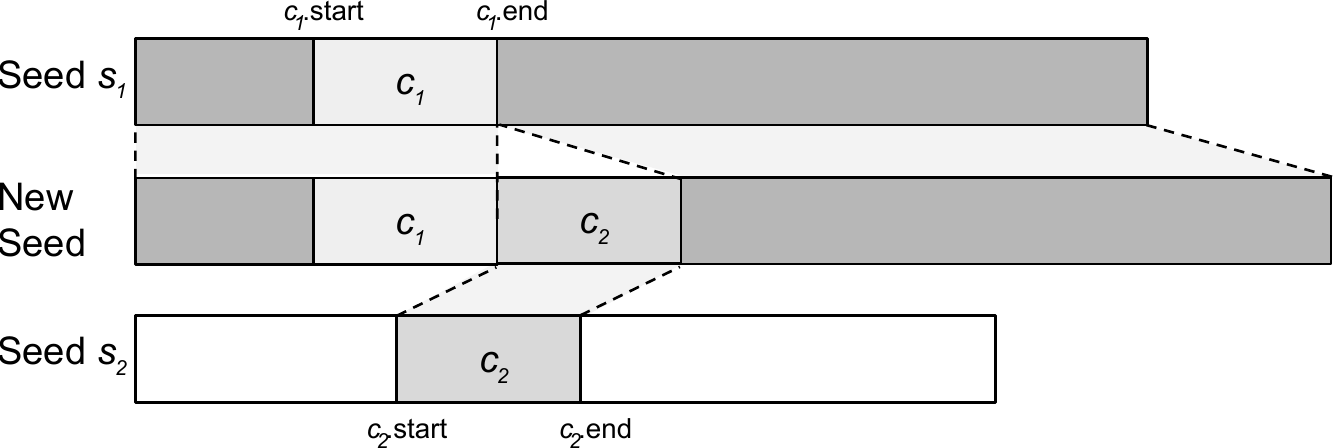}
\vspace{-0.2cm}
\end{figure}
\textbf{Smart Addition}.
Given a seed file $s_1$, choose an arbitrary second seed file $s_2$, choose an arbitrary chunk $c_2$ in $s_2$, and add it after an arbitrary existing chunk $c_1$ in $s_1$ that has a parent of the same type as $c_2$ (i.e., $c_1\text{\emph{.parent.type}} == c_2\text{\emph{.parent.type}}$).
The SGF copies the bytes following the end-index of $c_1$ to a new index where the length of the new chunk $c_2$ is added to the current end-index of the $c_1$ in the given seed file $s_1$. Then, the SGF copies the bytes between start- and end-index of $c_2$ in the second seed file $s_2$ to the end-index of the existing chunk $c_1$ in the given seed file $s_1$. Finally, all affected indices are revised in the virtual structure representing the generated input.

\begin{figure}[h]\centering
\vspace{-0.2cm}
\includegraphics[width=0.85\columnwidth]{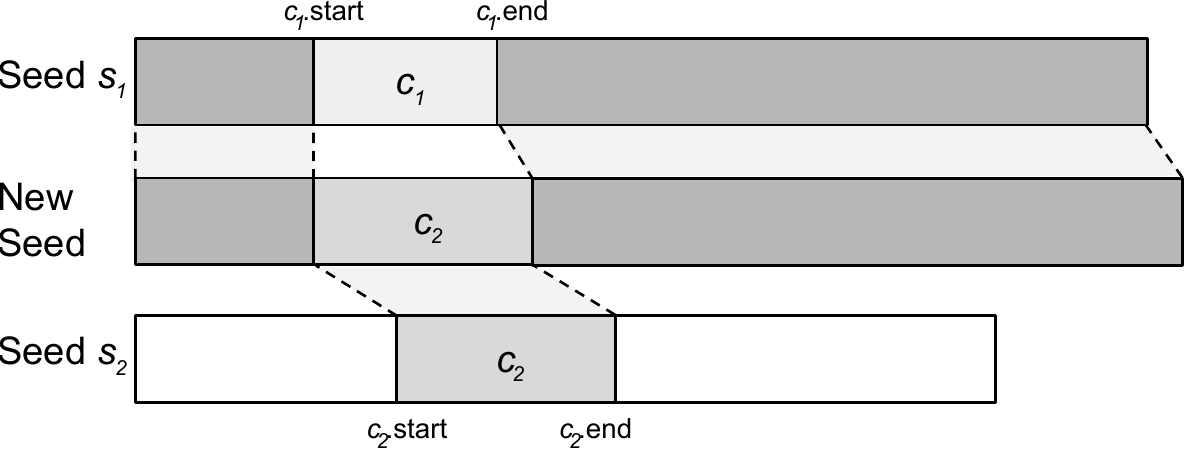}
\vspace{-0.2cm}
\end{figure}

\textbf{Smart Splicing}.
Given a seed file $s_1$, choose an arbitrary chunk $c_1$ in $s_1$, choose an arbitrary second seed file $s_2$, choose an arbitrary chunk $c_2$ in $s_2$ \emph{such that $c_1$ and $c_2$ have the same type}, and substitute $c_1$ with $c_2$.
The SGF copies the bytes following the end-index of $c_1$ to a new index where the length of the new chunk $c_2$ is added to the current end-index of the $c_1$ in the given seed file $s_1$. Then, the SGF copies the bytes between start- and end-index of $c_2$ in the second seed file $s_2$ to the end-index of the existing chunk $c_1$ in the given seed file $s_1$. Finally, all affected indices are revised in the virtual structure representing the generated input.

\textbf{Maintaining validity}. The files generated by applying structural mutation operators have a higher degree of validity than files generated by applying bit-level mutation operators. The specification of \emph{immutable attributes} allows the smart greybox fuzzer to apply bit-level mutation operators only to indices of \emph{mutable  attributes} (which are not structurally relevant), increasing the likelihood to generate valid files. However, there is no guarantee that our structural mutation operators maintain the validity of a file. For instance, in our motivating example the Peach Pit format specification may allow to add or delete \texttt{fmt} chunks while strictly speaking the formal WAVE format specification allows only exactly one \texttt{fmt} chunk. Nevertheless, it was our relaxed specification which allowed finding the vulnerability in the first place (it requires two \texttt{fmt} chunks to be present). In summary, \emph{strict} validity is not always desirable while a high degree of validity is necessary to reach beyond the parser code. This is a critical advantage of our lightweight  virtual structure design.

\subsection{Smart Mutation}
During \emph{smart mutation}, new inputs are generated by applying structural as well as simple mutation operators to the chosen seed file (cf. \textsc{mutate\_input} in Alg. \ref{alg:greybox}). In the following, we discuss the challenges and opportunities of smart mutation.

\subsubsection{Stacking Mutations}
To generate interesting test inputs, it might be worthwhile to apply several structural (high level) and bit-level (low level) mutation operators together. In mutation-based fuzzing, this is called \emph{stacking}. Bit-level mutation operators can easily be stacked in arbitrary order, knowing only the start- and end-index of the file. When data of length $x$ is deleted, we subtract $x$ from the end-index. When new data of length $x$ is added, we add $x$ to the new file's end-index.

However, it is not trivial to stack structural mutation operators. For each structural mutation, both the file itself and the virtual structure representing the file must be updated consistently. For instance, the deletion of a chunk will affect the end-indices of all its parent chunks, and the indices of every chunk ``to the right'' of the deleted chunk (i.e., chunks with a start-index that is greater than the deleted chunk's end-index).
Our implementation \smart makes a copy of the seed's virtual structure and stacks mutation operators by applying them consistently to both, the virtual structure and the file itself. This allows us to stack structural (high-level) mutation operators. Furthermore, if a bit-level (low-level) mutation operation cannot be translated into a mutation of the input structure, e.g., because bytes are deleted over chunk-boundaries, the mutation is not applied.

\subsubsection{Deferred Parsing}
\label{sec:deferred}
In our experiments, we observed that constructing the virtual structure for a seed input incurs substantial costs. The appeal of coverage-based greybox fuzzing (CGF) and the source of its success is its \emph{efficiency} \cite{boehme16coverage}. Generating and executing an input is in the order of a few milliseconds. However, we observed that parsing an input takes generally in the order of seconds. For instance, the construction of the virtual structure for a 218-byte PNG file takes between two and three seconds. If SGF constructs the virtual structure for every seed input that is discovered, SGF may quickly fall behind traditional greybox fuzzing despite all of its "smartness".

To overcome this scalability challenge, we developed a scheme that we call \emph{deferred parsing}, which contributed substantially to the scalability of our tool \smart.
We construct the virtual structure of a seed input with a certain probability $p$ that depends on the current time to discover a new path.
Let $t$ be the time since the last discovery of a new path. Let $s$ be the current seed chosen by \textsc{chooseNext} in Line~2 of greybox fuzzing Algorithm~\ref{alg:greybox} and assume that the virtual structure for $s$ has not been constructed, yet. Given a threshold $\epsilon$, we compute the probability $prob_{virtual}(s)$ to construct the virtual structure of $s$ as
\[
prob_{virtual}(s)=min\left(\frac{t}{\epsilon},1\right)\]
In other words, the probability $prob_{virtual}(s)$ to construct the virtual structure for the seed $s$ increases as the time $t$ since the last discovery increases.
Once $t \ge \epsilon$, we have $prob_{virtual}(s)=100$\%.

Our deferred parsing optimization is inspired by the following intuition. Without input aware greybox fuzzing as in \smart, AFL may generate many invalid inputs which repeatedly traverse a few short paths in an application
(typically program paths which lead to rejection of the input due to certain parse error). If more of such invalid inputs are generated, the value of $t$, the time since last discovery of a new path, is slated to increase.
Once $t$ increases beyond a threshold $\epsilon$, we allow \smart to construct the virtual structure. If however, normal AFL is managing to generate inputs which still traverse new paths, $t$ will remain small, and we will not incur the overhead of creating a virtual structure. The deferred parsing optimization thus allows \smart to achieve input format-awareness without sacrificing the efficiency of AFL.

\subsection{Validity-based Power Schedule}
A \emph{power schedule} determines how much energy is assigned to a given seed during coverage-based greybox fuzzing \cite{boehme16coverage}. The \emph{energy} for a seed determines how much time is spent fuzzing that seed when it is chosen next (cf. \textsc{assignEnergy} in Alg.~\ref{alg:greybox}). In the literature, several power schedules have been introduced. The original power schedule of AFL \cite{afl18} assigns more energy to smaller seeds with a lower execution time that have been discovered later. The gradient descent-based power schedule of \textsc{AFLfast} \cite{boehme16coverage} assigns more energy to seeds exercising low-frequency paths.

In the following, we define a simple validity-based power schedule. Conventionally, validity is considered as a boolean variable: Either a seed is valid, or it is not. However, we suggest to consider validity as a ratio: A file can be valid to a certain degree. The \emph{degree of validity} $v(s)$ of a seed $s$ is determined by the parser that constructs the virtual structure. If all of the file can be parsed successfully, the degree of validity $v(s)=100$\%. If only 65\% of $s$ can be parsed successfully, its validity $v(s)=65$\%. The virtual structure for a file that is partially valid is also only partially constructed. To this partial structure, one chunk is added that spans the unparsable remainder of the file.

Given the seed $s$, the \emph{validity-based power schedule} $p_v(s)$ assigns energy as follows
\begin{align}
p_v(s) &=
\begin{cases}
2p(s) &\text{ if } v(s) \ge 50\% \text{ and } p(s) \le \frac{U}{2}\\
p(s) & \text{ if } v(s) < 50\% \\
U & \text{ otherwise}
\end{cases}
\end{align}
where $p(s)$ is the energy assigned to $s$ by the traditional greybox fuzzer's (specifically AFL's) original power schedule and $U$ is a maximum energy that can be assigned by AFL. This power schedule implements a \emph{hill climbing meta-heuristic} that \emph{always} assigns twice the energy to a seed that is at least 50\% valid and has an original energy $p(s)$ that is at most half the maximum energy $U$.

The validity-based power schedule assigns more energy to seeds with a higher degree of validity. First, the utility of the structural mutation operators increases with the degree of validity. Secondly, the hope is that more valid inputs can be generated from already valid inputs. The validity-based power schedule implements a \emph{hill climbing meta-heuristic} where the search follows a gradient descent. A seed with a higher degree of validity will \emph{always} be assigned higher energy than a seed with a lower degree of validity.

\section{File Format Specification}
\label{sec:specs}

The quality of file format specifications is crucial to the effectiveness and efficiency of smart greybox fuzzing. However, manually constructing such high-quality specifications of highly-structured and complicated file formats is normally criticized as a time-consuming and error-prone task. In this work, we have done an extensive research on many popular file formats (e.g., document, video, audio, image, executable and network packet files) and found the key insights based on which users can write specifications in a systematic way. These key insights explain the common structures of file formats. On the other hand, they also show the correlations between the completeness \& preciseness of data models and the success of smart greybox fuzzing.

\subsection{Insight-1. Chunk inheritance} Most file formats are composed of data chunks which normally share a common structure. Like an abstract class in Java and other object-oriented programming languagues (e.g., C++ and C\#), to write an input specification we start by modelling a generic chunk containing attributes that are shared across all chunks in the file format. Then, we model the concrete chunks which inherit the attributes from the generic chunk. Hence, we only need to insert/modify chunk-specific attributes.

\hfill

\lstset{%
   		framexleftmargin=0mm,
   		frame=single,
   		numbers=none,
		resetmargins=true,
   		basicstyle=\ttfamily\selectfont\footnotesize,
   	  frameround= tttt,
   	  breaklines = true,
   	  tabsize=1,
   	  captionpos=b,
   	  language=C,
   	  morekeywords={Bug}}
\begin{lstlisting}[frame=single, caption=Generic Chunk Model, label=lst:wavegeneric]
<DataModel name="Chunk">
 <String name="ckID" length="4" padCharacter=" "/>
 <Number name="cksize" size="32">
  <Relation type="size" of="Data"/>
 </Number>
 <Blob name="Data"/>
 <Padding alignment="16"/>
</DataModel>
\end{lstlisting}

\lstset{%
   		framexleftmargin=0mm,
   		frame=single,
   		numbers=none,
		resetmargins=true,
   		basicstyle=\ttfamily\selectfont\footnotesize,
   	  frameround= tttt,
   	  breaklines = true,
   	  tabsize=1,
   	  captionpos=b,
   	  language=C,
   	  morekeywords={Bug}}
\begin{lstlisting}[frame=single, caption=Format Chunk Model, label=lst:wavefmt]
<DataModel name="ChunkFmt" ref="Chunk">
   <String name="ckID" value="fmt "  token="true"/>
   <Block name="Data">
      <Number name="wFormatTag" size="16"/>
      <Number name="nChannels" size="16"/>
      <Number name="nSampleRate" size="32"/>
      <Number name="nAvgBytesPerSec" size="32"/>
      <Number name="nBlockAlign" size="16" />
      <Number name="nBitsPerSample" size="16"/>
   </Block>
</DataModel>
\end{lstlisting}

Listing~\ref{lst:wavegeneric} and Listing~\ref{lst:wavefmt} show an example of how the chunk inheritance can be applied to the input specification of the WAVE audio file format. The generic chunk model in Listing~\ref{lst:wavegeneric} specifies that each chunk has its chunk identifier, chunk size and chunk data in which the chunk size constraints the actual length of the chunk data. Moreover, each chunk could have padded bytes at the end to make it word (2 bytes) aligned. Listing~\ref{lst:wavefmt} shows the model of a \emph{format} chunk, a specific data chunk in WAVE file, which inherits the chunk size and padding attributes from the generic chunk. It only models chunk-specific attributes like its string identifier and what are stored inside its data.

People normally have a big concern that they need to spend lots of time reading the standard specification of a file format (which can be hundreds of pages long) to understand this high-level hierarchical chunks structure. However, we find that there exist Hex editor tools like 010Editor~
\cite{editor010} which can detect the file format and quickly decompose a sample input file into chunks with all attributes. The tool currently supports 114 most common file formats (e.g., PDF, MPEG4, AVI, ZIP, JPEG)~\cite{templates010}.

\begin{figure}[h]
\includegraphics[width=\columnwidth]{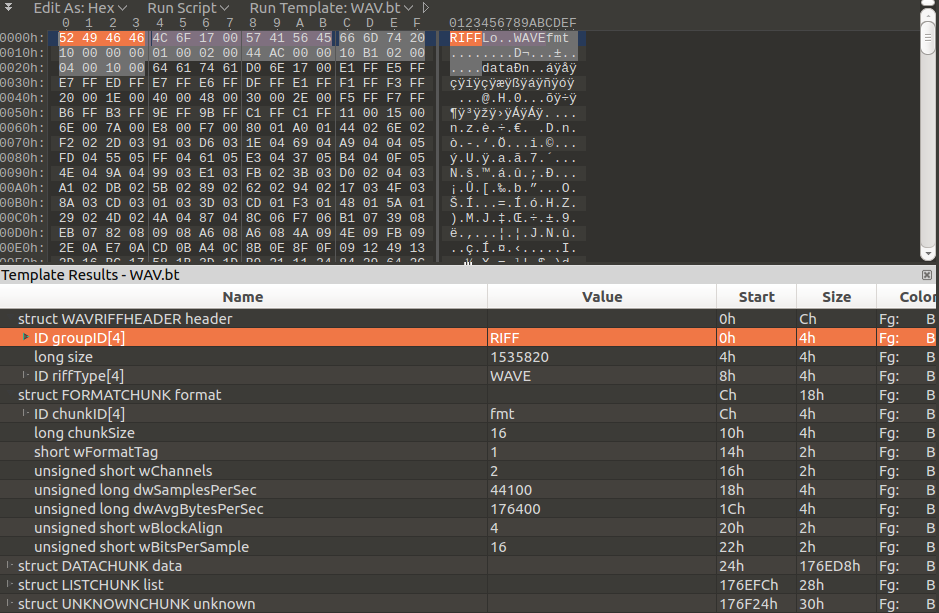}
\caption{Analyzing file structure using 010Editor}
\label{fig:editor}
\end{figure}

Figure~\ref{fig:editor} is a screenshot of 010Editor displaying a WAVE file. The top part of the screen shows the raw data in both Hexadecimal and ASCII modes. The bottom part is the decomposed components including chunks' headers, and chunks' data.

\subsection{Insight-2. Specification completeness} As explained in Section~\ref{sec:grammar}, smart greybox fuzzing supports structural mutation operators that work at chunk level. So we are not required to specify all attributes inside a chunk. We can start with a coarse-grained specification and gradually make it more complete. Listing~\ref{lst:wavefmtsimple} shows a simplified definition of the format chunk in which we only specify the chunk identifier and we do not define what are the children attributtes in its data. The chunk data is considered as a ``blob'' which can contain anything as long as its size is consistent with the chunk size.

\hfill

\lstset{%
   		framexleftmargin=0mm,
   		frame=single,
   		numbers=none,
		resetmargins=true,
   		basicstyle=\ttfamily\selectfont\footnotesize,
   	  frameround= tttt,
   	  breaklines = true,
   	  tabsize=1,
   	  captionpos=b,
   	  language=C,
   	  morekeywords={Bug}}
\begin{lstlisting}[frame=single, caption=Simplified Format Chunk Model, label=lst:wavefmtsimple]
<DataModel name="ChunkFmt" ref="Chunk">
   <String name="ckID" value="fmt "  token="true"/>
</DataModel>
\end{lstlisting}

Based on the this key insight and the Insight-1, one can quickly write a short yet precise file format specification. As shown in Section~\ref{sec:setup}, the specification for the WAVE file format can be written in 82 lines while the specification for the PCAP network traffic file format can be written in just 24 lines. These two specifications helped smart greybox fuzzing discover many vulnerabilities which could not be found by other baseline techniques.

\subsection{Insight-3. Relaxed constraints} There could be many constraints in a chunk (e.g., the chunk identifier must be a constant string, the chunk size attribute must match with the actual size or chunks must be in order). However, since the main goal of fuzzing or stress testing in general is to explore corner cases, we should relax some constraints as long as these relaxed constraints do not prevent the parser from decomposing the file. Listing~\ref{lst:wavfull} shows the definition of a WAVE file format. As we use the \emph{Choice} element\footnote{In a Peach pit, Choice elements are used to indicate any of the sub-elements are valid but only one should be selected at a time. Reference: \url{http://community.peachfuzzer.com/v3/Choice.html}} to specify the list of potential chunks (including both mandatory and optional ones), many constraints have been relaxed. Firstly, the chunks can appear in any order. Secondly, some chunk (including mandatory chunk) can be absent. Thirdly, some unknown chunk can appear. Lastly, some chunk can appear more than once. In fact, becaused this relaxed model, vulnerabilities like the one in our motivating example in our paper (Section \ref{sec:motivating}) can be exposed.

\hfill

\lstset{%
   		framexleftmargin=0mm,
   		frame=single,
   		numbers=none,
		resetmargins=true,
   		basicstyle=\ttfamily\selectfont\footnotesize,
   	  frameround= tttt,
   	  breaklines = true,
   	  tabsize=1,
   	  captionpos=b,
   	  language=C,
   	  morekeywords={Bug}}
\begin{lstlisting}[frame=single, caption=WAVE File Format Specification, label=lst:wavfull]
<DataModel name="Wav">
  <String name="ckID" value="RIFF" token="true"/>
  <Number name="cksize" size="32" />
  <String name="WAVE" value="WAVE" token="true"/>
  <Choice name="Chunks" maxOccurs="30000">
    <Block name="FmtChunk" ref="ChunkFmt"/>
    <Block name="DataChunk" ref="ChunkData"/>
    <Block name="FactChunk" ref="ChunkFact"/>
    <Block name="SintChunk" ref="ChunkSint"/>
    <Block name="WavlChunk" ref="ChunkWavl"/>
    <Block name="CueChunk" ref="ChunkCue"/>
    <Block name="PlstChunk" ref="ChunkPlst"/>
    <Block name="LtxtChunk" ref="ChunkLtxt"/>
    <Block name="SmplChunk" ref="ChunkSmpl"/>
    <Block name="InstChunk" ref="ChunkInst"/>
    <Block name="OtherChunk" ref="Chunk"/>
  </Choice>
</DataModel>
\end{lstlisting}

\subsection{Insight-4. Reusability} Unlike specifications of program behaviours which are program specific and hardly reusable, a file format specification can be used to fuzz all programs taking the same file format. We believe the benefit of finding new vulnerabilities far outweighs the cost of writing input specifications. In Section~\ref{sec:setup} and Section~\ref{sec:results}, we show that our smart greybox fuzzing tool have used specifications of 10 popular file formats (PDF, AVI, MP3, WAV, JPEG, JPEG2000, PNG, GIF, PCAP, ELF) to discover more than 40 vulnerabilities in heavily-fuzzed real-world software packages. Notably, based on the key insights we have presented, it took one of us only five (5) working days to complete these 10 specifications.

\section{Experimental Setup}
\label{sec:setup}
To evaluate the effectiveness and efficiency of smart greybox fuzzing, we conducted several experiments. We implemented our technique by extending the existing greybox fuzzer AFL and call our smart greybox fuzzer \smart. To investigate whether input-structure-awareness indeed improves the vulnerability finding capability of a greybox fuzzer, we compare \smart with two traditional greybox fuzzers AFL \cite{afl18} and \fast \cite{boehme16coverage}. To investigate whether smart blackbox fuzzer (given the same input model) could achieve a similar vulnerability finding capability, we compare \smart with the smart blackbox fuzzer Peach \cite{peach18}. We also compare \smart with \vuzzer~\cite{rawat17vuzzer}. The objective of \vuzzer is similar to \smart, it seeks to tackle the challenges of structured file formats for greybox fuzzing, yet without input specifications, using taint analysis and control flow analysis.

\subsection{Research Questions}
\begin{itemize}
  \item[\textbf{RQ-1.}] \emph{Is smart greybox fuzzing more effective and efficient than traditional greybox fuzzing?} Specifically, we investigate whether \smart exposes more unique crashes than AFL/\fast in 24 hours, and in the absence of crashes whether \smart explores more paths than AFL/\fast in the given time budget.
  \item[\textbf{RQ-2.}] \emph{Is smart greybox fuzzing more effective and efficient than smart blackbox fuzzing?} Specifically, we investigate whether \smart exposes more unique crashes than Peach in 24 hours, and in the absence of crashes whether \smart explores more paths than Peach in the given time budget.
  \item[\textbf{RQ-3.}] \emph{Is smart greybox fuzzing more effective than taint analysis-based greybox fuzzing?}  Specifically, we investigate the number of bugs found by each technique individually and all together.

\end{itemize}

\begin{table*}\footnotesize\centering
\caption{Subject Programs and File Formats. \vuzzer subjects are at the bottom.}
\begin{tabular}{llr|lrl}
\hline
\cline{1-6}
Program          & Description              & Size (LOC)    & Test driver & Format     & Option \\
\hline
Binutils        & Binary analysis utilities	& 3700 K 		& \texttt{readelf}        & \textbf{ELF}       &  \texttt{-agteSdcWw --dyn-syms -D @@} \\
Binutils        & Binary analysis utilities	& 3700 K 		& \texttt{nm-new}      & \textbf{ELF}      & \texttt{-a -C -l --synthetic @@} \\
LibPNG        & Image processing	& 111 K		& \texttt{pngimage}       & \textbf{PNG} 	& \texttt{@@} \\
ImageMagick      & Image processing	 & 385 K		 & \texttt{magick}        & \textbf{PNG} 	& \texttt{@@ /dev/null} \\
LibJPEG-turbo        & Image processing 		& 87 K	& \texttt{djpeg}        & \textbf{JPEG} 	& \texttt{@@} \\
LibJasper        & Image processing		& 33 K	& \texttt{imginfo}        & \textbf{JPEG} 	    & \texttt{-f @@} \\
FFmpeg        & Video/Audio/Image processing		& 1100 K	& \texttt{ffmpeg}        & \textbf{AVI} 	&  \texttt{-y -i @@ -c:v mpeg4 -c:a out.mp4}        \\
LibAV        & Video/Audio/Image processing		& 670 K	& \texttt{avconv}        & \textbf{AVI} 	  & \texttt{-y -i @@ -f null -}       \\
LibAV        & Video/Audio/Image processing		& 670 K	& \texttt{avconv}        & \textbf{WAV} 	  & \texttt{-y -i @@ -f null -}       \\
WavPack        & Lossless Wave file compressor		& 47 K	& \texttt{wavpack}        & \textbf{WAV} 	   &  \texttt{-y @@ -o out\_dir}     \\
OpenJPEG        & Image processing		& 115 K	& \texttt{decompress}        & \textbf{JP2} 	   &  \texttt{-y @@ -o out\_dir}     \\
LibJasper        & Image processing		& 33 K	& \texttt{jasper}        & \textbf{JP2} 	   &  \texttt{-y @@ -o out\_dir}     \\
\hline
\hline
mpg321        & Command line MP3 player		& 5 K	& \texttt{mpg321}        & \textbf{MP3} 	   &  \texttt{--stdout @@}     \\
gif2png+libpng        & Image converter		& 36 K	& \texttt{gif2png}        & \textbf{GIF} 	   &  \texttt{@@}     \\
pdf2svg+libpoppler        & PDF to SVG converter		& 92 K	& \texttt{pdf2svg}        & \textbf{PDF} 	   &  \texttt{@@ out.svg}     \\
tcpdump+libpcap        & Network traffic analysis		& 102 K	& \texttt{tcpdump}        & \textbf{PCAP} 	   &  \texttt{-nr @@}     \\
tcptrace+libpcap        & TCP connection analysis		& 55 K	& \texttt{tcptrace}        & \textbf{PCAP} 	   &  \texttt{@@}     \\
djpeg+libjpeg        & Image processing 			& 37 K	& \texttt{djpeg}        & \textbf{JPEG} 	& \texttt{@@} \\
\hline
\end{tabular}
\label{tab:subjects}
\end{table*}

\subsection{Implementation: \smart}
\begin{figure}[h]\centering
\includegraphics[width=\columnwidth]{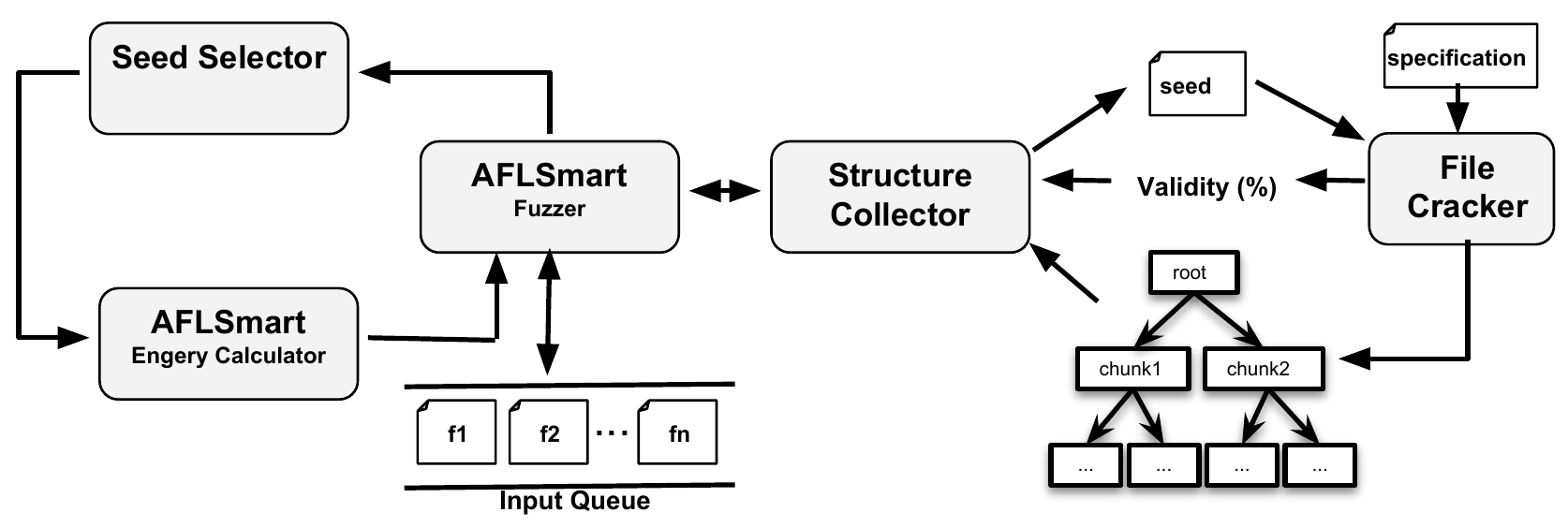}
\caption{Architecture of \smart}
\label{fig:implementation}
\end{figure}

\smart extends AFL by adding and modifying four components, the File Cracker, the Structure Collector, the Energy Calculator and the Fuzzer itself. The overall architecture is shown in Figure~\ref{fig:implementation}. While currently integrated with Peach, we designed \smart such that it provides a general framework that allows integrating other input parsers and to define further structural mutation operators.

{\bf{\smart File Cracker}} parses an input file and decomposes it into data chunks and data attributes. It also calculates the validity of the input file based on how much of the file can be parsed. In this prototype, we implement the File Cracker by modifying the Cracker component of the smart blackbox fuzzer Peach (Community version) \cite{peach18} which fully supports highly-structured file formats such as PNG, JPEG, GIF, MP3, WAV and AVI.

{\bf{\smart Structure Collector}} connects the core \smart Fuzzer and the File Cracker component. When the Fuzzer requests structure information of the current input to support its operations (e.g., smart mutations), it passes the input to the Structure Collector for collecting the validity and the decomposed chunks and attributes. This component provides a generic interface to support all File Crackers -- our current Peach-based File Cracker and new ones. It is also worth noting that \smart Fuzzer only collects these information once and saves them for future uses.

{\bf{\smart Energy Calculator}} implements the validity-based power schedule as discussed in Section \ref{sec:grammar}. Hence, \smart assigns more energy to inputs which are more syntactically valid. Specifically, we apply a new formula to the \emph{calculate\_score} function of \smart.

{\bf{\smart Fuzzer}} contains the most critical changes to make \smart effective. In this component, we design and implement the virtual structure which can represent input formats in a hierarchical structure. Based on this core data structure, all \smart mutation operations which work at chunk levels are implemented. We also modify the \emph{fuzz\_one} function of AFL to support our important optimizations -- deferred parsing and stacking mutations (Section~\ref{sec:grammar}).

Note that our changes do not impact the instrumentation component of AFL. As a result, we can use \smart to fuzz program binaries provided the binary is instrumented using a tool like DynamoRio \cite{dynamorio} and the instrumented code can be processed by AFL. Such a binary fuzzing approach has been achieved in the WinAFL tool\footnote{https://github.com/ivanfratric/winafl} for Windows binaries. \smart works well with such binary fuzzing tools.

\subsection{Subject Programs}

We did a rigorous search for suitable benchmarks to test \smart and the chosen baselines. We evaluated the techniques using both large real-world software packages and a benchmark previously used in \vuzzer paper. We did not use the popular LAVA benchmarks~\cite{sp16lava} because the LAVA-M subjects (\emph{uniq, base64, md5sum, who}) do not process structured files while the small \emph{file} utility in LAVA-1 takes any file, regardless of its file format, and determines the file type.

In the comparison with AFL, \fast and Peach (RQ-1 and RQ-2), we selected the newest versions (at the time of our experiments) of 11 experimental subjects from well-known open source programs which take ten (6) highly-structured file formats -- executable binary file (ELF), image files (PNG, JPEG, JP2 (JPEG2000)), audio/video files (WAV, AVI). All of them have been well tested for many years. Notably, five (5) media processing libraries (FFmpeg\footnote{https://github.com/FFmpeg/FFmpeg}, LibPNG\footnote{https://github.com/glennrp/libpng}, LibJpeg-Turbo\footnote{https://github.com/libjpeg-turbo}, ImageMagick\footnote{https://github.com/ImageMagick/ImageMagick}, and OpenJPEG\footnote{https://github.com/uclouvain/openjpeg}) have joined the Google OSS-Fuzz project\footnote{\url{https://github.com/google/oss-fuzz}} and they are continuously tested using the state-of-the-art fuzzers including AFL and LibFuzzer. LibAV\footnote{https://github.com/libav/libav}, WavPack \footnote{https://github.com/dbry/WavPack} and Libjasper\footnote{https://github.com/mdadams/jasper} are widely-used libraries and tools for image, audio and video files processing and streaming. Binutils\footnote{https://www.gnu.org/software/binutils/} is a set of utilities for analyzing binary executable files. It is installed on almost all Linux-based machines.

To compare with \vuzzer (RQ-3), we chose the same benchmark used in the paper. The benchmark includes old versions of six (6) popular programs on Ubuntu 14.04 32-bit: mpg321 (v0.3.2), gif2png (v2.5.8), pdf2svg (v0.2.2), tcpdump (v4.5.1), tcptrace (v6.6.7), and djpeg (v1.3.0). These subjects take MP3, GIF, PDF, PCAP and JPEG files as inputs. It is worth noting that \vuzzer has not supported 64-bit environment yet.

Table~\ref{tab:subjects} shows the full list of programs and their information. Note that the sizes of subject programs are calculated by \texttt{sloccount}.\footnote{https://www.dwheeler.com/sloccount/}. Moreover, to increase the reproducibility of our experiments, we also provide the exact command options we used to run the subject programs. In the experiments to answer RQ-1 and RQ-2, we tested two programs for each file format to mitigate subject bias.

\subsection{Corpora, Dictionaries, and Specifications}
\emph{Format specification}.
\smart leverages  file format specifications to construct the virtual structure of a file. These specifications are developed as Peach Pits.\footnote{\url{http://community.peachfuzzer.com/v3/PeachPit.html}} In our experiment, we used ten file format specifications (see Table \ref{tab:models}). While the specification of the WAV format is a modification of a free Peach sample\footnote{\url{http://community.peachfuzzer.com/v3/TutorialFileFuzzing/}}, we developed other Peach pits from scratch. \smart and Peach are provided with the same file format specifications (i.e., Peach pits).

\emph{Seed corpus}. In order to construct the initial seed files, we leveraged several sources. For PNG and JPEG images, we used the image files that are available as test files in their respective code repositories. For ELF files, we collected program binaries from the \emph{bin} and \emph{/user/bin} folders on the host machine. For other file formats, we downloaded seed inputs from websites keeping sample files (WAV\footnote{\url{https://freewavesamples.com/source/roland-jv-2080}}, AVI\footnote{\url{http://www.engr.colostate.edu/me/facil/dynamics/avis.htm}}, JP2\footnote{http://samples.ffmpeg.org/}, PCAP\footnote{https://wiki.wireshark.org/SampleCaptures}, MP3\footnote{https://www.magnac.com/sounds.shtml}, GIF\footnote{https://people.sc.fsu.edu/~jburkardt/data/gif/gif.html} and PDF\footnote{https://www.pdfa.org/isartor-test-suite/}). Table~\ref{tab:models} shows the size of the input corpus we used for each file format. All fuzzers are provided with the same initial seed corpus.


\begin{table}\small
\caption{File Format Specifications and Seed Corpora}
\label{tab:models}
{\small
\begin{tabular}{lrl|rr}
\hline
\multicolumn{3}{c|}{File Format Specification} & \multicolumn{2}{c}{Seed Corpus}\\
Format          & Length (\#Lines)              & Time spent    & \#Files & Avg. size      \\
\hline
ELF        	& 90 lines        	& 4 hours             & 21 		& 100 KB                   \\
PNG        	& 128 lines	       & 4 hours             & 51	 	& 4 KB                     \\
JPEG        	& 92 lines	       & 4 hours             & 8	 	& 5.5 KB                     \\
WAV        	& 82 lines	       & 1 hour             & 11	 	& 500 KB                     \\
AVI        	& 124 lines	       & 4 hours             & 10	 	& 430 KB                     \\
JP2        	& 144 lines	       & 4 hours             & 10	 	& 35 KB                     \\
PDF        	& 84 lines	       & 4 hours             & 10	 	& 140 KB                     \\
GIF        	& 108 lines	       & 4 hours             & 10	 	& 12 KB                     \\
PCAP        	& 24 lines	       & 4 hours             & 5	 	& 11 KB                     \\
MP3        	& 90 lines	       & 4 hours             & 10	 	& 201 KB                     \\
\hline
\end{tabular}}
\vspace{-0.2cm}
\end{table}

\emph{Dictionary}. We developed dictionaries for four (4) file formats (ELF, WAV, AVI, and JP2); AFL (and \smart) already provides dictionaries for PNG and JPEG image formats. The dictionaries were written by simply crafting the tokens (e.g., signatures, chunk types) from the same specifications/documents based on which we developed the Peach Pit file format specifications. Both \smart and AFL were run with dictionaries.

\emph{Reproducibility}. To ensure the reproducibility of our experiments, we will make \smart open source and provide the seed corpora, dictionaries, and Peach Pits used.

\subsection{Infrastructure}

\emph{Computational Resources}. We have different setups for two sets of experiments. In the first set of experiments to compare \smart with AFL, \fast, and Peach we used machines with an Intel Xeon CPU E5-2660v3 processor that has 56 logical cores running at 2.4GhZ. Each machine runs Ubuntu 16.04 (64 bit) and has access to 64GB of main memory. All fuzzers have the same time budget (24 hours), the same computational resources, and are started with the same seed corpus with the same dictionaries. Peach and \smart also use the same Peach Pits.

In the comparison with \vuzzer, as \vuzzer has not supported 64-bit environment yet, we set up a virtual machine (VM) having the same settings reported in the paper -- a Ubuntu 14.04 LTS system equipped with a 32-bit 2-core Intel CPU and 4 GB RAM. Both \vuzzer and \smart are started with the same seed corpus.

\emph{Experiment repetition}. To mitigate the impact of randomness, for each subject program we run five (5) isolated instances of each of AFL, \fast, \smart, and Peach in parallel. We emphasize that none of the instances share the same queue. Specifically, Peach does not support the shared queue architecture (i.e., parallel fuzzing mode in AFL\footnote{\url{https://github.com/mirrorer/afl/blob/master/docs/parallel_fuzzing.txt}}).

\emph{Measurement in AFL-based fuzzers}. The greybox fuzzers AFL, \fast, and \smart already provide the number of explored paths in five-second intervals in a file called \texttt{plot\_data}. This allows us to plot these quantities over time. To compute the number of unique bugs found, we used a call stack-based bucketing approach \cite{rebucket} to analyze and group the discovered bugs. Crashes that have the exact the same call stack are in the same group. We selected one representative from each group for bug reporting purposes.

\emph{Measurement in Peach}.
Peach does not keep the generated test cases. It only stores bug-triggering inputs which complicates our measurement of the number of paths explored. Hence, we modified Peach such that we could collect all test cases which Peach generates during a 24-hour run. Then, we use the \emph{afl-cmin}\footnote{\url{https://github.com/mirrorer/afl/blob/master/afl-cmin}} -- a corpus minimization utility in the AFL toolset to find the smallest subset of files in the generated test cases that still trigger the full range of instrumentation data points. To achieve a fair comparison, we also use the same \emph{afl-cmin} to minimize the test cases generated by AFL, \fast and \smart. These results are reported in the fourth column (\#Min-set) of the Table~\ref{tab:results}



\section{Experimental Results}\vspace{-0.2cm}
\label{sec:results}

\subsection*{RQ.1\quad SGF Versus Traditional Greybox Fuzzing}

\begin{table}[ht!]\footnotesize
\caption{Average number of paths discovered, the minimal sets of test cases calculated by afl-cmin, crashes found, and unique bugs discovered in 5 runs after 24 hours.}
\label{tab:results}
\begin{tabular}{l|lrrrc}
\hline
\cline{1-6}
\scriptsize {Binary}          & \scriptsize {Fuzzer}              & \scriptsize {\#Paths}    & \scriptsize {\#Min-set} 	& \scriptsize {\#Crashes}  & \scriptsize {\#Bugs}    \\
\hline
\texttt{readelf}       	& AFL    		  & 14855    	& 6285	& 15         & 3     \\
ELF		   	& \fast           	  & 16048      	& 6422 	& 22          & 3               \\
		   		& Peach           	& N/A      		& 1202 	& 0             & 0    \\
                			& \smart\           & 16236   & 7002   & 19             & 3           \\
\cline{2-6}
 \texttt{nm-new}            	& AFL    		  & 10201     & 4283	& 33                  & 1                \\
ELF		   	& \fast          & 10159     & 3995	& 45                           & 1   \\
			& Peach           & N/A      	& 454	& 0                       & 0         \\
                		& \smart\           & 8981        & 3885     & 34  	    & 2             \\
\hline
\texttt{pngimage}       	& AFL    		 & 5280    & 2324	 & 0          & 0                        \\
PNG		   	& \fast          & 5663     & 2294		& 0                        & 0       \\
			& Peach           & N/A     & 	395 	& 0                         & 0         \\
                		& \smart\         		& 6497 	& 2560    & 1                & 1                  \\
\cline{2-6}
\texttt{magick}       	& AFL    		  & 6434     & 2696	& 0                      & 0        \\
PNG		   	& \fast          & 6249     & 2668	& 0                        & 0      \\
			& Peach           & N/A      & 66	& 0                        & 0         \\
                		& \smart\           & 6860   & 2861  & 0             & 0                     \\
\hline
\texttt{djpeg}      		& AFL    		  & 3661		& 1275     & 0                & 0                   \\
JPEG		   	& \fast           & 3778     & 1264	& 0                          & 0         \\
			& Peach           & N/A     & 342		& 0                             & 0     \\
               		& \smart\           & 4005     & 1351		& 0                            & 0    \\
\cline{2-6}
\texttt{imginfo}       	& AFL    		 & 1681     &	967		& 18                           & 2       \\
JPEG		   	& \fast          & 1437    & 759		 & 44                            & 2       \\
			& Peach           & N/A     & 53		 & 0                              & 0     \\
                		& \smart\           & 1812    & 1003	 & 58                          & 2       \\
\hline
\texttt{ffmpeg}       	& AFL    		  & 2783   & 1340   & 0              & 0                  \\
AVI	   	& \fast           & 3378     & 1547	 & 0                     & 0            \\
			& Peach           & N/A    & 1413	  & 0                    & 0            \\
                		& \smart\           & 8485     & 3582		 & 2             & 1                    \\
\cline{2-6}
\texttt{avconv}       	& AFL    		 & 4980     & 1205	& 213              & 3                    \\
AVI	   	& \fast          & 4900    & 1209	 & 218                     & 3            \\
			& Peach           & N/A     & 849		 & 0                       & 0           \\
                		& \smart\          & 13549    & 3328	 & 503                      & 3             \\
\hline
\texttt{avconv}       & AFL    		  & 14849   & 4271  & 0               & 0                    \\
WAV		   	& \fast           & 14617    & 4209	 & 0                        & 0          \\
			& Peach           & N/A   & 867	   & 0                           & 0      \\
                		& \smart\          & 20616    & 6418	 & 13                  & 3               \\
\cline{2-6}
\texttt{wavpack}       	& AFL    		  & 1724   & 425  & 59                    & 1               \\
WAV		   	& \fast           & 1950  & 460   & 48                        & 1          \\
			& Peach           & N/A   & 339	   & 0                          & 0       \\
                		& \smart\           & 1998   & 537	  & 191                     & 5              \\
\hline
\texttt{decompress}       & AFL    		  	& 6615   & 1984  & 0               & 0                    \\
JPEG2000		   	& \fast           		& 6767    & 2030	 & 0                        & 0          \\
					& Peach           		& N/A   & 389	   & 0                           & 0      \\
                		& \smart\          			& 6503    & 1950	 & 16                  & 3               \\
\cline{2-6}
\texttt{jasper}       	& AFL    		  & 2624   & 1049  & 220                    & 6               \\
JPEG2000		   	& \fast           & 2298  & 954   & 156                        & 5          \\
			& Peach          		 & N/A   & 215	   & 0                          & 0       \\
                		& \smart\           & 3957   & 1582	  & 944                     & 10              \\
\hline
\end{tabular}
\end{table}

\begin{table}\footnotesize
\caption{Bug reports. Assertion Failure (AF), Aborted (AB), Divide-by-Zero (DZ), Heap/Stack Overflow (OF), Null Pointer Reference (NP)}
\label{tab:bugreport}
\begin{tabular}{@{}l@{ \ }l@{\ }l@{ \ }c@{ \ }c@{ \ }c@{ \ }c@{}}
\hline
\cline{1-7}
\scriptsize {Subject}          & \scriptsize {Bug-ID}              & \scriptsize {Type}    			& \scriptsize {AFL}     & \scriptsize {\fast} 	& \scriptsize {Peach} 		& \scriptsize {\smart}     \\
\hline
WavPack		&CVE-2018-10536        & OF    		 & \xmark      & \xmark 	 & \xmark        & \cmark                     \\
			&CVE-2018-10537        & OF    		 & \xmark      & \xmark 	 & \xmark        & \cmark                     \\
			&CVE-2018-10538        & OF    		 & \xmark      & \xmark 	 & \xmark        & \cmark                     \\
			&CVE-2018-10539       & OF    		 & \xmark      & \xmark 	 & \xmark        & \cmark                     \\
			&CVE-2018-10540        & OF    		 & \cmark      & \cmark 	 & \xmark        & \cmark                     \\
\cline{1-7}
Binutils		&Bugzilla-23062       & AF    		 & \cmark       & \cmark 	 & \xmark       & \cmark                     \\
			&Bugzilla-23063       & AF    		 & \cmark       & \cmark 	 & \xmark       & \cmark                     \\
			&CVE-2018-10372       & OF    		 & \cmark       & \cmark 	 & \xmark     & \cmark                     \\
			&CVE-2018-10373        & NP    		 & \cmark       & \cmark 	 & \xmark     & \cmark                     \\
			&Bugzilla-23177      & OF    		 & \xmark       & \xmark 	 & \xmark      & \cmark                     \\
\cline{1-7}
LibPNG		&CVE-2018-13785       & DZ    		 	& \xmark    & \xmark 	 & \xmark          & \cmark                     \\
\cline{1-7}
Libjasper	&Issue-174       & AF    		& \cmark        & \cmark 	 & \xmark      & \cmark                     \\
			&Issue-175       & AF    		& \cmark             & \cmark 	 & \xmark 	   & \cmark                     \\
			&Issue-182-1       & OF    		& \xmark             & \xmark 	 & \xmark 	   & \cmark                     \\
			&Issue-182-2       & NP    		& \xmark             & \xmark 	 & \xmark 	   & \cmark                     \\
			&Issue-182-3       & OF    		& \xmark             & \xmark 	 & \xmark 	   & \cmark                     \\
			&Issue-182-4       & NP    		& \xmark             & \xmark 	 & \xmark 	   & \cmark                     \\
			&Issue-182-5       & OF    		& \cmark             & \cmark 	 & \xmark 	   & \cmark                     \\
			&Issue-182-6       & AF    		& \cmark             & \cmark 	 & \xmark 	   & \cmark                     \\
			&Issue-182-7       & AF    		& \cmark             & \cmark 	 & \xmark 	   & \cmark                     \\
			&Issue-182-8       & AB    		& \cmark             & \cmark 	 & \xmark 	   & \cmark                     \\
			&Issue-182-9       & AF    		& \cmark             & \cmark 	 & \xmark 	   & \cmark                     \\
			&Issue-182-10       & AF    		& \cmark             & \xmark 	 & \xmark 	   & \cmark                     \\
\cline{1-7}
OpenJPEG	&Email-Report-1       & OF    		 & \xmark      & \xmark 	 & \xmark        & \cmark                     \\
			&Email-Report-2       & OF  		 & \xmark      & \xmark 	 & \xmark        & \cmark                     \\
			&Issue-1125       & AF    		 & \xmark      & \xmark 	 & \xmark        & \cmark                     \\
\cline{1-7}
LibAV		&Bugzilla-1121       & OF    		 & \xmark      & \xmark 	 & \xmark        & \cmark                     \\
			&Bugzilla-1122       & OF  		 & \xmark      & \xmark 	 & \xmark        & \cmark                     \\
			&Bugzilla-1123       & OF    		 & \xmark      & \xmark 	 & \xmark        & \cmark                     \\
			&Bugzilla-1124       & OF    		 & \cmark      & \cmark 	 & \xmark        & \cmark                     \\
			&Bugzilla-1125       & DZ	    		 & \cmark      & \cmark 	 & \xmark        & \cmark                     \\
			&Bugzilla-1127        & OF	    		 & \cmark      & \cmark 	 & \xmark        & \cmark                     \\
\cline{1-7}
FFmpeg		&Email-Report-3       	 & DZ   		 & \xmark      & \xmark 	 & \xmark        & \cmark                     \\
\hline
\hline
\textbf{TOTAL} 	&      		&   		 & 16      & 15 	 & 0        & 33                     \\
\hline
\end{tabular}
\end{table}

\begin{figure*}
\includegraphics[width=\textwidth]{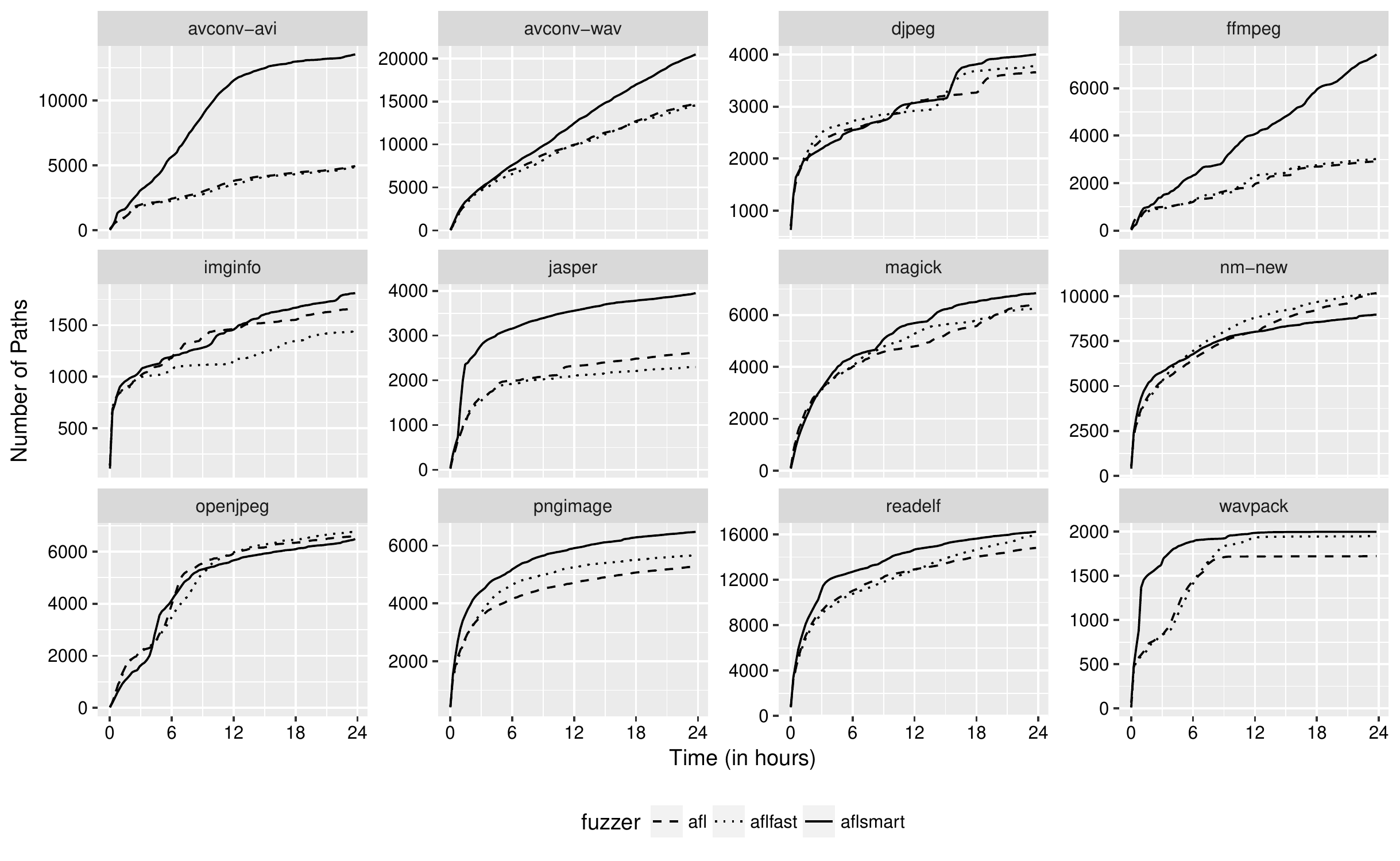}
\caption{Number of paths discovered over time for AFL, \fast, and \smart}
\label{fig:paths}
\end{figure*}

\emph{In terms of discovered number of paths, \smart clearly outperforms both AFL and \fast.}
\smart discovered more paths in ten (10) out of twelve (12) subjects. In the two larger subjects, ffmpeg and avconv (taking AVI files), \smart explored 200\% more paths than AFL and \fast. The same improvement can be observed in the minimized sets of test cases (\#Min-set) as well. \smart  performed a bit worse than AFL and \fast (in terms of path exploration) in a ELF-parsing subject in Binutils (nm-new) and an OpenJPEG utility (decompress). For these two subjects, \smart achieved similar path coverage in the first six (6) hours after which AFL and \fast started outperforming \smart (see Figure~\ref{fig:paths}).

\emph{In terms of bug finding, \smart discovered bugs in 10 subjects while AFL and \fast could not detect bug in four of them (ffmpeg, pngimage, decompress and avconv (taking WAV files))}. After analyzing the crashes, we reported 33 zero-day bugs found by \smart out of which only 16 bugs were found by AFL and \fast. Vice versa, all zero-day bugs that AFL and \fast found were also found by \smart. Hence, \smart discovered twice as many bugs as AFL/\fast. Table~\ref{tab:bugreport} shows the detailed bugs found by \smart and the baseline. 17 bugs are heap \& stack buffer overflows (many of them are buffer overwrites) which are known to be easily exploitable. The maintainers of these programs have fixed 12 bugs we reported. The MITRE corporation\footnote{\url{https://cve.mitre.org/}} has assigned eight (8) CVEs to the most critical vulnerabilities.

The main reason why AFL and \fast could not find many bugs, meanwhile \smart did, in subjects like FFmpeg, LibAV, WavPack, and OpenJPEG is that these programs take in highly structured media files (e.g., image, audio, video) in which the data chunks must be placed in order at correct locations. This is very challenging for traditional greybox fuzzing tools like AFL and \smart. In addition to the motivating example (CVE-2018-10536 and CVE-2018-10537), we analyze in depth few more critical vulnerabilities found by \smart to explain the challenges.

\textbf{CVE-2018-10538: Heap Buffer Overwrite}. The \emph{buffer overwrite} is caused by two integer overflows and insufficient memory allocation. To construct an exploit, we need to craft a \emph{valid WAVE file} that contains the mandatory \texttt{riff}, \texttt{fmt}, and \texttt{data} chunks. Between the \texttt{fmt} and \texttt{data} chunk, we add an \emph{additional unknown chunk} (i.e., that is neither fmt, data, ..) with \texttt{cksize} $ \ge 0\text{x}80000000$.

\begin{figure}[h]\footnotesize
\begin{tabular}{@{}l@{\quad}l}
\tiny{286}&\footnotesize \texttt{\textbf{else} \{}\quad\quad\quad\quad\quad\quad\emph{// just copy unknown chunks to output file}\\
\tiny{287}&\\
\tiny{288}&\footnotesize \texttt{\textbf{int} bytes\_to\_copy=(chunk\_header.ckSize+1) \& ~1L;}\\
\tiny{289}&\footnotesize \texttt{\textbf{char} *buff=malloc(bytes\_to\_copy);}\\
\ldots\\
\tiny{296}&\footnotesize \texttt{\textbf{if} (!DoReadFile(infile,buff,bytes\_to\_copy,..)) \{}\\
\end{tabular}
\caption{Showing \texttt{cli/riff.c} @ revision \texttt{0a72951}}
\label{fig:vuln3}
\end{figure}

During parsing the file, WavPack enters the ``unknown chunk'' handling code shown in Figure~\ref{fig:vuln3}. It reads the specified chunk size from the \texttt{chunk\_header} struct and stores it as a 32-bit \emph{signed} integer. Since \texttt{ckSize} $ \ge 2^{31}$, the assignment in \texttt{riff.c:288} overflows, such that \texttt{bytes\_to\_copy} contains a negative value. The memory allocation function \texttt{malloc} takes only unsigned values causing a second overflow to a smaller positive number. When \texttt{DoReadFile} attempts to read more information from the WAVE file, there is not enough memory being allocated, resulting in a memory overwrite that can be controlled by the attacker.
This vulnerability (CVE-2018-10538) was patched by aborting when \texttt{bytes\_to\_copy} is negative.

\textbf{OpenJPEG-1: Heap Buffer Overread \& Overwrite}.

The buffer overread (lines 617-619) and overwrite (lines 629-631) (see Figure~\ref{fig:vuln4}) are caused by a missing check of the actual size (width and height) of the three color streams (red, green, and blue). Without this check, the code assumes that all the three streams have the same size and it uses the same bound value (\emph{max}) to access the buffers. To construct an exploit, we need to craft a \emph{valid} JP2 (JPEG2000) file that contains three color streams having different sizes by ``swapping'' the whole stream(s) from one valid JP2 file and place it/them in the correct position(s) in another valid JP2 file. Without the structural information, traditional greybox fuzzing is unlikely to do such a precise swapping.

\begin{figure}[h]\footnotesize
\begin{tabular}{@{}l@{\quad}l}
\tiny{612}&\footnotesize \texttt{r = image->comps[0].data;}\\
\tiny{613}&\footnotesize \texttt{g = image->comps[1].data;}\\
\tiny{614}&\footnotesize \texttt{b = image->comps[2].data;}\\
\ldots\\
\tiny{616}&\footnotesize \texttt{for (i = 0U; i < max; ++i) \{}\\
\tiny{617}&\footnotesize \quad \texttt{*in++ = (unsigned char) * r++;}\\
\tiny{618}&\footnotesize \quad \texttt{*in++ = (unsigned char) * g++;}\\
\tiny{619}&\footnotesize \quad \texttt{*in++ = (unsigned char) * b++;}\\
\tiny{620}&\footnotesize \texttt{\}}\\
\ldots\\
\tiny{622}&\footnotesize \texttt{cmsDoTransform(transform, inbuf, outbuf, ...);}\\
\ldots\\
\tiny{624}&\footnotesize \texttt{r = image->comps[0].data;}\\
\tiny{625}&\footnotesize \texttt{g = image->comps[1].data;}\\
\tiny{626}&\footnotesize \texttt{b = image->comps[2].data;}\\
\ldots\\
\tiny{628}&\footnotesize \texttt{for (i = 0U; i < max; ++i) \{}\\
\tiny{629}&\footnotesize \quad \texttt{*r++ = (unsigned char) * out++;}\\
\tiny{630}&\footnotesize \quad \texttt{*g++ = (unsigned char) * out++;}\\
\tiny{631}&\footnotesize \quad \texttt{*b++ = (unsigned char) * out++;}\\
\tiny{632}&\footnotesize \texttt{\}}\\
\end{tabular}
\caption{Showing \texttt{common/color.c} @ rev \texttt{d2205ba}}
\label{fig:vuln4}
\end{figure}

\vspace{0.2cm}
\subsection*{RQ.2\quad SGF Versus Smart Blackbox Fuzzing}
\emph{Given the same input format specifications, \smart clearly outperforms Peach in all twelve (12) subjects} (see Table~\ref{tab:results} and Table~\ref{tab:bugreport}). \smart generated up to an order of magnitude meaningful test cases (see \#Min-set column in Table~\ref{tab:results}) and discovered 33 zero-day bugs while Peach could not find a single vulnerability .\footnote{Unlike for the AFL-based fuzzers, Peach does not produce data that allows us to plot the number of paths discovered over time in Figure~\ref{fig:paths}.}

Apart from the difficulty to discover zero-day bugs in the heavily-fuzzed benchmarks, we explain these results by the lack of coverage feedback mechanism in Peach. The smart blackbox fuzzer treats all test cases at all stages equally. There is no evolution of a seed corpus. Instead, there is a simple enumeration of files that are valid w.r.t. the provided specification. This is a well-kown limitation of Peach. Recently Lian et. al \cite{peachcov} have tried to tackle this problem by applying LLVM passes and designing a feedback mechanism for Peach. The tool is not available for further comparison and analysis.

A second explanation is the completeness of the file format specification. The performance of Peach substantially depends on the precision and completeness of the file format specification. Peach might need more detailed input models in which (almost) all chunks and attributes are specified with exact data types to generate more interesting files. In contrast, \smart does not require very detailed file format specifications to derive the virtual structure of a file and apply our structural mutation operators.

\subsection*{RQ.3\ \ Versus Taint analysis-based Greybox Fuzzing}

\emph{\smart outperforms \vuzzer on a \vuzzer's benchmark}. \smart found 15 bugs in all subject programs in the benchmark in which seven (7) bugs could not be found by \vuzzer in \emph{tcpdump}, \emph{tcptrace} and \emph{gif2png} (see Table~\ref{tab:Vuzzers}. It is worth noting that all these bugs are not zero-day ones because the \vuzzer benchmark contains old versions of software packages on the out-dated Ubuntu 14.04 32-bit; all the bugs have been fixed. We explain these results by the limited information \vuzzer can infer using taint analysis -- it cannot infer the high-level structural representation of the input so it cannot do mutations at the chunk level.

\begin{table}\small
\caption{\vuzzer vs \smart on \vuzzer's benchmark}
\label{tab:Vuzzers}
{\small
\begin{tabular}{l|rl|rl}
\hline
\multicolumn{1}{c|}{Application} & \multicolumn{2}{c|}{Vuzzer} & \multicolumn{2}{c}{\smart}\\
          					& \#Crashes       & \#Bugs     & \#Crashes       & \#Bugs      \\
\hline
mpg321        			& 337        	& 2             & 193 		& 2                   \\
gif2png+libpng        		& 127	       & 1             & 54	 		& 2                     \\
pdf2svg+libpoppler        	& 13	       & 3             & 20	 		& 2                     \\
tcpdump+libpcap        	& 3	       	& 1             & 149	 	& 6                     \\
tcptrace+libpcap        		& 403	       & 1             & 240	 	& 2                     \\
djpeg+libjpeg        		& 1	      	 	& 1             & 1	 		& 1                     \\
\hline
\end{tabular}}
\vspace{-0.2cm}
\end{table}


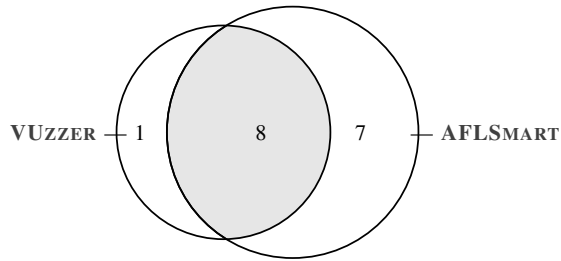
\begin{figure}[t]\centering
\vspace{-0.5cm}
\def\secondcircle{(0:-0.25cm) circle (1.7cm)}
\def\thirdcircle{(0:0.85cm) circle (2.0cm)}
\colorlet{circle edge}{black}
\colorlet{circle area}{gray!20}
\tikzset{filled/.style={fill=circle area, draw=circle edge, thick},
outline/.style={draw=circle edge, thick}}
\setlength{\parskip}{5mm}
\resizebox{0.9\columnwidth}{!}{
\begin{tikzpicture}
\begin{scope}
    \clip \secondcircle;
    \fill[filled] \thirdcircle;
\end{scope}
\draw[outline] \secondcircle; \node [draw=none] at (180:0.10) {\textbf{\color{darkgray}{\vuzzer}} ---\ 1\quad\quad\quad\quad\quad \quad\quad\quad\quad\quad\quad\quad\quad\quad\quad};
\draw[outline] \thirdcircle; \node [draw=none] at (180:-2.5) {\quad 8\quad\quad\quad\quad 7\quad\quad --- \textbf{\color{darkgray}{\smart}}};
\end{tikzpicture}
}
\caption{Venn Diagram showing the number of bugs that \vuzzer and \smart discover individually and together.}
\label{fig:micro-venn}
\end{figure}

We also investigate the intersection of the results. As shown in Figure~\ref{fig:micro-venn}, \vuzzer and \smart discovered 16 bugs all together. Even though the intersection is large (\smart discovered almost all bugs found by \vuzzer), we believe \smart and \vuzzer are two potentially supplementary approaches. While \smart can leverage the input structure information to systematically do mutations at the chunk level and explore new search space (which is unlikely to be done by bit-level mutations), \vuzzer can leverage its taint analysis to infer features of attributes inside the newly generated inputs and mutate them effectively.

\section{Case Study. Bug Hunting using \smart}

We conducted an extra experiment to evaluate the effectiveness of \smart in a bug hunting campaign for a large and popular software package. We chose FFmpeg as our target program because this is an extremely popular and heavily-fuzzed library. Every day when we use our computers/smartphones in working time or in our leisure time, we would use at least one software powered by the FFmpeg library like a web browser (e.g., Google Chrome), a sharing video page (e.g., YouTube), or a media player (e.g., VLC). FFmpeg is heavily fuzzed; as a part of OSS-Fuzz project, it has been continuously fuzzed for years. Due to its popularity, any serious vulnerability in FFmpeg could compromise millions of systems and expose critical security risk(s).

We run five (5) instances of \smart in parallel mode\footnote{\url{https://github.com/mirrorer/afl/blob/master/docs/parallel_fuzzing.txt}} in one week using the AVI input specification to test its functionality of converting an AVI file to a MPEG4 file (see Table~\ref{tab:subjects} for the exact command). In this fuzzing campaign, \smart discovered nine (9) zero-day crashing bugs including buffer overflows, null pointer dereferences and assertion failures. All the bugs have been fixed and nine (9) CVE IDs have been assigned to them. Table~\ref{tab:ffmpegbug} shows the CVEs and their severity levels based on the Common Vulnerability Scoring System version 3.0\cite{cvss3}; all these nine vulnerabilities are rated from medium to high severity.

\begin{table}\centering
\caption{CVEs of bugs found in FFmpeg}
\label{tab:ffmpegbug}
\begin{tabular}{@{}llll@{}}
\hline
\cline{1-4}
Subject          & Bug-ID              & Description    	 & Severity		\\
\hline
FFmpeg		&CVE-2018-13301       & Null pointer dereference    	& MEDIUM	  \\
			&CVE-2018-13305        & Heap buffer overwrite    	& HIGH	  \\
			&CVE-2018-13300        & Heap buffer overread    	& HIGH	  \\
			&CVE-2018-13303       & Null pointer dereference    	& MEDIUM	  \\
			&CVE-2018-13302        & Heap buffer overwrite    	& HIGH	  \\
			&CVE-2018-12459        & Assertion failure    		& MEDIUM  \\
			&CVE-2018-12458        & Assertion failure    		& MEDIUM  \\
			&CVE-2018-13304        & Assertion failure    		& MEDIUM  \\
			&CVE-2018-12460        & Null pointer dereference    	& MEDIUM	  \\
\hline
\end{tabular}
\vspace*{-0.1in}
\end{table}

The results confirm the practical impact of smart greybox fuzzing in testing programs taking highly-structured input files like FFmpeg. It shows that the benefit of finding new vulnerabilities outweighs the one-time effort of writing input specifications.

\section{Related Work}
\label{sec:related}

\emph{Smart blackbox fuzzing}. The stream of works that is most closely related to ours is that of smart blackbox fuzzers which leverage file format specifications to generate inputs for a program that is otherwise treated as a blackbox.
In the area of smart blackbox fuzzing, input grammars have been used to generate
test inputs~\cite{purdom72sentence}. There exist a
variety of tools employing this technique, such as Peach
fuzzer~\cite{peach18}, Spike~\cite{aitel18spike},
Domato~\cite{fratric18domato}, and LangFuzz~\cite{holler12fuzzing}. LangFuzz is a
smart blackbox fuzzer that has been used to detect crashes in JavaScript engines; it uses a file format specification to mutate a given seed input and replaces code fragments with those learned from a set of parsed sample inputs. Our work on \smart can be seen as integrating the format-awareness capability into coverage-based grey-box fuzzing.

\emph{Smart whitebox fuzzing}. Another related stream of works is that of smart whitebox fuzzing which leverages both program structure and input structure to explore the program most effectively. Whitebox fuzzers are often based on symbolic execution engines such as KLEE~\cite{cadar08klee}, or S$^2$E~\cite{chipounov11s2e}.
Grammar-based whitebox fuzzers \cite{godefroid08grammar} can generate files that are valid w.r.t. a context-free grammar.
Model-based whitebox fuzzing \cite{pham16model} enforces semantic constraints over the input structure that cannot be expressed in a context-free grammar, such as length-of relationships.
In contrast to our approach, smart whitebox fuzzers require heavy machinery of symbolic execution and constraint solving.

\emph{Coverage-based greybox fuzzing}.
Our work builds on coverage-based greybox fuzzing
(CGF)~\cite{afl18,libfuzzer18}, which is a popular and effective
approach for software vulnerability detection. The AFL fuzzer
\cite{afl18} and its extensions \cite{boehme16coverage,boehme17aflgo,li17steelix,peng18fuzzing,stephens16driller,lemieux17fairfuzz,chen18angora,sp18collafl}
constitute the most widely used embodiment of CGF. CGF is a promising
middle ground between blackbox and whitebox fuzzing.  Compared to
blackbox approaches, CGF uses light-weight instrumentation to guide
the fuzzer to new regions of the code, and compared to whitebox
approaches, CGF does not suffer from high overheads of constraint
solving \cite{efficiency}. To the best of our knowledge, ours is the first work to propose and
build an input format-aware greybox fuzzer.

\emph{Boosted greybox fuzzing}.
\textsc{AFLfast}~\cite{boehme16coverage} uses
Markov chain modeling to target regions that are still not generally
covered by AFL. The approach discovers known bugs faster compared to
standard AFL, as well as finding new bugs. \textsc{AFLgo}
\cite{boehme17aflgo} performs reachability analysis to a given location or target
by prioritizing seeds which are estimated to have a lower distance to the target.
Angora \cite{chen18angora} is an extension of AFL to improve its coverage
that performs search based on gradient descent to solve path condition
without symbolic execution. SlowFuzz~\cite{petsios17slowfuzz}
prioritizes inputs with a higher resource usage count for further
mutation, with the objective of discovering vulnerabilities to complexity attacks.
These works improve the effectiveness of greybox fuzzing along other dimensions (not input
format awareness), and are largely orthogonal to our approach

\emph{Restricted mutations}.
Other works in the CGF area employ specific optimizations to restrict the mutations.
VUzzer~\cite{rawat17vuzzer} uses
data- and control-flow analysis of the test subject to detect the locations and
the type of the input data to mutate or to keep constant.
Steelix~\cite{li17steelix} focuses on developing customized mutation operations of \emph{magic
  bytes}, e.g., the special words \texttt{RIFF}, \texttt{fmt}, or
\texttt{data} in a WAVE file (see \ref{sec:motivating}). SymFuzz~\cite{cha15program} learns the dependencies in the bits in the
seed input using symbolic execution in order to compute an optimal \emph{mutation
ratio\/} given a program under test and the seed input; the mutation ratio is the number of seed bits
that are flipped in mutation-based fuzzing.
These works encompass specific optimizations to restrict mutations. They do {\em not} inject input format awareness for
generating valid inputs as is achieved by our file format aware mutation operators, or validity-based power schedules.

\emph{Greybox fuzzing and symbolic execution}.
T-Fuzz~\cite{peng18fuzzing} removes sanity checks in the
code that blocks the fuzzers (AFL or honggfuzz~\cite{honggfuzz18})
from progressing further. This, however, introduces false
positives, which are then detected using symbolic execution.
Driller~\cite{stephens16driller} is a combination of fuzzing  and symbolic execution to allow for
deep exploration of program paths.
In our work, we avoid any symbolic execution, and enhance the effectiveness of grey-box fuzzing without sacrificing the efficiency of AFL.

\emph{Format specification inference}.
Several works study file format inferencing.
Lin and Zhang \cite{lin08deriving} present an approach to derive the file's input
tree from the dynamic execution trace.
Learn\&Fuzz~\cite{godefroid17learn} uses neural-network-based
statistical machine learning to generate files satisfying a complex
format. The approach is used to fuzz Microsoft Edge browser PDF
handler, and found a bug not previously found by previous approaches
such as SAGE~\cite{godefroid12sage}. \textsc{Autogram} \cite{autogram} uses dynamic taint analysis to derive input grammars. Such works on input format inference can
potentially help input-aware fuzzers such as \smart.

\section{Discussion}
\label{sec:discussion}

Greybox fuzzing has been the technology of choice for practical, automated detection of software vulnerabilities. The current embodiment of greybox fuzzing in the form of the AFL fuzzer is agnostic to the input format specification. This leads to lot of time in a fuzzing campaign being wasted in generation of syntactically invalid inputs. In this work, we have brought in the input format awareness of commercial blackbox fuzzers into the domain of greybox fuzzing. This is achieved via file format aware mutations, validity-based power schedules, and several optimizations (most notably the deferred parsing optimization) which allows our \smart tool to retain the efficiency of AFL. Detailed evaluation of our tool \smart with respect to AFL on applications processing popular file formats (such as AVI, MP3, WAV) demonstrate that \smart achieves substantially (up to 200\%) higher path coverage and finds more bugs as compared to AFL. The manual effort of specifying an input format is a one-time effort, and was limited to 4 hours for each of the input formats we examined.

In future, we can extend the input file-format fuzzing of \smart to input protocol fuzzing by taking into account input protocol specifications, along the lines of the state model already supported by the Peach fuzzer. This will allow us to extend \smart for fuzzing of reactive systems. Moreover, the recent work of Godefroid et al. \cite{godefroid17learn} has shown the promise of learning input formats automatically, albeit for a specific format namely PDF. In future, we plan to study this direction to further alleviate the one-time manual effort of specifying an input format. Last but not the least, we can use the flexible architecture of \smart (Figure~\ref{fig:implementation}) to support interfacing with many other input-format-aware blackbox fuzzers, such as the Domato fuzzer \cite{fratric18domato} which is known to work well for HTML format.
This will enhance the utility of \smart for a wider variety of file formats.

\section*{Acknowledgments}
This research was partially supported
by a grant from the National Research Foundation, Prime
Minister’s Office, Singapore under its National Cybersecurity
R\&D Program (TSUNAMi project, No. NRF2014NCRNCR001-21)
and administered by the National Cybersecurity
R\&D Directorate.

\end{document}